\documentclass[11pt]{article}
\usepackage[dvipsnames,svgnames,x11names]{xcolor}
\usepackage{titlesec}
\usepackage[T1]{fontenc}
\usepackage{tikz}
\usetikzlibrary{shapes.geometric,arrows}
\tikzset{
  basics/.style={minimum width=30mm, minimum height=7.5mm, text centered, draw=black},
  basicsone/.style={minimum width=15mm, minimum height=15mm, text centered, draw=black},
  startstop/.style={rectangle, rounded corners, basics, fill=red!30},
  io/.style={trapezium, trapezium left angle=70, trapezium right angle=110, basics, fill=blue!30},
  process/.style={rectangle, basics},
  processone/.style={rectangle, basics},
  processtwo/.style={rectangle, basicsone},
  decision/.style={ellipse,basics},
  arrow/.style={thick,->,>=stealth},
  connected/.style={dashed,-},
}

\usepackage{amsmath, amsthm}
\usepackage{newpxtext}
\usepackage{newpxmath}

\usepackage{fullpage}

\usepackage[colorlinks = true]{hyperref}
\hypersetup{
 linkcolor=[rgb]{0.3,0.3,0.6},
 citecolor=[rgb]{0.2, 0.6, 0.2},
 urlcolor=[rgb]{0.6, 0.2, 0.2}
}
\usepackage[capitalise, noabbrev, nameinlink]{cleveref}

\usepackage{mathtools}
\usepackage{todonotes}
\usepackage{tikz}
\usetikzlibrary{decorations.pathreplacing,backgrounds}
\usepackage{multirow}

\usepackage{stackrel}
\usepackage{algorithm}
\usepackage{algpseudocode}
\usepackage{comment}
\usepackage{algorithm}
\usepackage{algpseudocode}
\usepackage{accents}

\usepackage{todonotes}

\usepackage{bm}

\usepackage{fancybox}
\usepackage{color}
\usepackage{latexsym}

\usepackage{eepic}
\usepackage{epic}
\usepackage{epsf}
\usepackage{graphics}
\usepackage{dsfont}
\usepackage{anyfontsize}

\usepackage{graphicx} 

\newtheorem{theorem}{Theorem}
\newtheorem*{theorem*}{Theorem}

\newtheorem{lemma}[theorem]{Lemma}

\newtheorem{fact}[theorem]{Fact}
\newtheorem{corollary}[theorem]{Corollary}
\newtheorem{claim}[theorem]{Claim}
\newtheorem{observation}[theorem]{Observation}
\newtheorem{proposition}[theorem]{Proposition}
\theoremstyle{definition}
\newtheorem{definition}[theorem]{Definition}
\newtheorem{remark}[theorem]{Remark}
\newenvironment{claimproof}[1]{\par\noindent{\em Proof of Claim.}~#1}

\newcommand{\B}{\mathfrak{B}}

\newcommand{\M}{\mathrm{Mat}}
\renewcommand{\mod}{\textrm{mod}}
\renewcommand{\H}{\mathcal{H}}

\newcommand{\F}{\mathds{F}}
\newcommand{\G}{\mathcal{G}}

\newcommand{\Q}{\mathds{Q}}

\newcommand{\rit}{\rm\text{RIT}}

\DeclareMathOperator{\poly}{poly}
\DeclareMathOperator{\rank}{rank}
\DeclareMathOperator{\Span}{span}
\DeclareMathOperator{\Gal}{Gal}

\DeclareMathOperator{\sing}{\rm{\textsc{Singular}}}
\DeclareMathOperator{\nsing}{\rm{\textsc{NSingular}}}

\newcommand{\Mat}{\mbox{\small\rm Mat}} 
\newcommand{\p}{{\mathscr{p}}}

\newcommand{\NC}{\mbox{\small\rm NC}}

\renewcommand{\ge}{\geqslant}
\renewcommand{\geq}{\geqslant}
\renewcommand{\le}{\leqslant}
\renewcommand{\leq}{\leqslant}

\DeclareFontFamily{OMX}{MnSymbolE}{}
\DeclareFontShape{OMX}{MnSymbolE}{m}{n}{
   <-6>  MnSymbolE5
   <6-7>  MnSymbolE6
   <7-8>  MnSymbolE7
   <8-9>  MnSymbolE8
   <9-10> MnSymbolE9
   <10-12> MnSymbolE10
   <12->   MnSymbolE12}{}
\DeclareSymbolFont{mnlargesymbols}{OMX}{MnSymbolE}{m}{n}
\SetSymbolFont{mnlargesymbols}{bold}{OMX}{MnSymbolE}{b}{n}
\DeclareMathDelimiter{\llangle}{\mathopen}{mnlargesymbols}{'164}{mnlargesymbols}{'164}
\DeclareMathDelimiter{\rrangle}{\mathclose}{mnlargesymbols}{'171}{mnlargesymbols}{'171}

\renewcommand{\angle}[1]{\langle #1 \rangle}
\newcommand{\ubar}[1]{\underaccent{\bar} #1}
\newcommand{\dangle}[1]{{\llangle} #1 {\rrangle}}

\DeclareFontEncoding{LS1}{}{}
\DeclareFontSubstitution{LS1}{stix}{m}{n}
\DeclareSymbolFont{symbols2stix}{LS1}{stixfrak} {m} {n}
\DeclareMathSymbol{\lparenless}{\mathopen} {symbols2stix}{"32}
\DeclareMathSymbol{\rparengtr}{\mathclose}{symbols2stix}{"33}

\DeclareMathOperator{\ord}{ord}

\newcommand{\newbrak}[1]{{\lparenless} #1 {\rparengtr}}

\newcommand{\n}{\mathfrak{n}}

\title{Black-Box Identity Testing of Noncommutative Rational Formulas in Deterministic Quasipolynomial Time\footnote{A preliminary version of this work appeared in the proceedings of the \textit{56th Annual ACM Symposium on Theory of Computing (STOC, 2024)}.}}
\author{
V. Arvind\thanks{Institute of Mathematical Sciences (HBNI), and 
Chennai Mathematical Institute, Chennai, India. \texttt{Email: arvind@imsc.res.in.}} \and
Abhranil Chatterjee\thanks{Indian Institute of Technology, Kanpur, India. \texttt{Email: abhneil@gmail.com}.} 
\and Partha Mukhopadhyay\thanks{Chennai Mathematical Institute, Chennai, India. \texttt{Email: partham@cmi.ac.in}.}
}

\date{}
\begin{document}
\maketitle
\begin{abstract}
Rational Identity Testing ($\rit$) is the decision problem of determining whether or not a noncommutative rational formula computes zero in the free skew field. It admits a  deterministic polynomial-time white-box algorithm
\cite{GGOW16, IQS18, HH21}, and a randomized polynomial-time algorithm~\cite{DM17} in the black-box setting, via singularity testing of linear matrices over the free skew field. Indeed, a randomized $\NC$ algorithm for $\rit$ in the white-box setting follows from the result of Derksen and Makam \cite{DM17}. 

Designing an efficient deterministic black-box algorithm for $\rit$ and understanding the parallel complexity of $\rit$ are major open problems in this area. Despite being open since the work of Garg, Gurvits, Oliveira, and Wigderson \cite{GGOW16}, these questions have seen limited progress. In fact, the only known result in this direction is the construction of a quasipolynomial-size hitting set for rational formulas of only \emph{inversion height} two \cite{ACM22}. 

In this paper, we significantly improve the black-box complexity of this problem and obtain the first quasipolynomial-size hitting set for \emph{all} rational formulas of polynomial size. 
Our construction also yields the first deterministic quasi-$\NC$ upper bound for $\rit$ in the white-box setting.
\end{abstract}

\newpage
\tableofcontents

\newpage
\section{Introduction}\label{sec:intro}
The goal of algebraic circuit complexity is to understand the
complexity of computing multivariate polynomials and rational
expressions using basic arithmetic operations, such as additions,
multiplications, and inverses. Algebraic formulas and algebraic
circuits are some of the well-studied computational models.

In the commutative setting, the role of inverses is well understood,
but in noncommutative computation it is quite subtle. To elaborate, it
is known that \emph{any} commutative rational expression can be
expressed as $fg^{-1}$ where $f$ and $g$ are two commutative
polynomials~\cite{Str73}. However, noncommutative rational expression
such as $x^{-1} + y^{-1}$ or $xy^{-1}x$ cannot be represented as $f
g^{-1}$ or $f^{-1} g$ for any noncommutative polynomials $f$ and
$g$. Therefore, the presence of \emph{nested inverses} makes a
rational expression more complicated, for example ${(z +
  xy^{-1}x)}^{-1}- z^{-1}$.

A noncommutative rational expression is not always defined on a matrix
substitution. For a noncommutative rational expression $\Phi$, its
\emph{domain of definition} is the set of matrix tuples (of any
dimension) where $\Phi$ is defined.

Two rational expressions $\Phi_1$ and $\Phi_2$ are \emph{equivalent}
if they agree on every matrix substitution in the intersection of
their domains of definition. This induces an equivalence relation on
the set of all noncommutative rational expressions (with nonempty
domains of definition). Interestingly, this computational definition
was used by Amitsur in the characterization of the \emph{universal}
free skew field \cite{Amitsur66}. The free skew field consists of
these equivalence classes, called \emph{noncommutative rational
functions}. One can think of the free skew field $\F\newbrak{x_1,
  \ldots, x_n}$ as the smallest field that contains the noncommutative
polynomial ring $\F\angle{x_1, \ldots, x_n}$. The free skew field been
extensively studied in mathematics~\cite{Amitsur66, Coh71, Coh95,
  FR04}.

The complexity-theoretic study of noncommutative rational functions
was initiated by Hrube\v{s} and Wigderson \cite{HW15}.
Computationally (and in this paper), noncommutative rational functions
are represented by algebraic formulas using addition, multiplication,
and inverse gates over a set of noncommuting variables, and they are
called noncommutative rational formulas.  Hrube\v{s} and Wigderson
\cite{HW15} also addressed the \emph{rational identity testing}
problem ($\rit$): decide efficiently whether a given noncommutative
rational formula $\Phi$ computes the zero function in the free skew
field. Equivalently, the problem is to decide whether $\Phi$ is zero
on its domain of definition, follows from Amitsur's characterization
\cite{Amitsur66}.

For example, the rational expression $(x+xy^{-1}x)^{-1} + (x+y)^{-1}
-x^{-1}$ is a rational identity, known as Hua's identity~\cite{Hua49}.
Rational expressions exhibit peculiar properties which seem to make
the $\rit$ problem quite different from the noncommutative polynomial
identity testing. For example, Bergman has constructed an explicit
rational formula, of inversion height two, which is an identity for
$3\times 3$ matrices but not an identity for $2\times 2$ matrices
\cite{Breg76}.  Also, the apparent lack of \emph{canonical
representations}, like a sum of monomials representation for
polynomials, and the use of nested inverses in noncommutative rational
expressions complicate the problem.  This motivates the definition of
\emph{inversion height} of a rational formula which is the maximum
number of inverse gates in a path from an input gate to the output
gate.  The \emph{inversion height} of a rational function is the
minimum over all the inversion heights of formulas representing the
function. For example, consider the rational expression
$(x+xy^{-1}x)^{-1}$. Athough it has a nested inverse, by Hua's
identity it represents a rational function of inversion height one. In
fact, Hrube\v{s} and Wigderson obtain the following interesting bound
on the inversion height of any rational function \cite{HW15}. This is
obtained by adapting Brent's depth reduction for the commutative
formulas \cite{Brent74}.

\begin{fact}\label{fact:inv-height-bound}
For any noncommutative $n$-variate rational formula $\Phi_1$ of size
$s$, one can construct a rational formula $\Phi_2$ of size $s$ with
the following properties:
\begin{enumerate}
\item Both $\Phi_1$ and $\Phi_2$ compute the same rational function.
\item The domains of definition of $\Phi_1$ and $\Phi_2$ are identical.
\item The inversion height of $\Phi_2$ is at most $O(\log s)$.
\end{enumerate}

\end{fact}

Consequently, to design a black-box RIT algorithm for rational
formulas of size at most $s$, it suffices to construct a hitting set
for rational formulas of inversion height at most $O(\log s)$. This
upper bound is crucial for our proof.

Hrube\v{s} and Wigderson have given an efficient reduction from the
$\rit$ problem to the singularity testing problem of linear matrices
in noncommuting variables over the free skew field ($\nsing$).
Equivalently, given a linear matrix $T = A_1x_1 + \ldots+ A_nx_n$ over
noncommuting variables $\{x_1, x_2, \ldots, x_n\}$, the problem
$\nsing$ is to decide whether or not there is a matrix substitution
$(p_1, \ldots, p_n)$ such that $\det \left( \sum_{i=1}^n A_i\otimes
p_i\right)\neq 0$ \cite{IQS18}.  It is the noncommutative analogue of
Edmonds' problem of symbolic determinant identity testing ($\sing$).
While $\sing$ can be easily solved in randomized polynomial time using
Polynomial Identity Lemma \cite{DL78, Zip79, Sch80}, finding a
deterministic polynomial-time algorithm remains completely
elusive~\cite{KI04}.

Remarkably, $\nsing\in \textrm{P}$ thanks to two independent
breakthrough results \cite{GGOW16, IQS18}.  In particular, the
algorithm of Garg, Gurvits, Oliveira, and Wigderson \cite{GGOW16} is
analytic in nature and based on operator scaling which works over
$\Q$. The algorithm of Ivanyos, Qiao, and Subrahmanyam \cite{IQS18} is
purely algebraic.  Moreover, the algorithm in their paper~\cite{IQS18}
works over both $\Q$ and fields of positive
characteristic. Subsequently, a third algorithm based on convex
optimization was developed by Hamada and Hirai \cite{HH21}. Not only
are these beautiful results, but also they have enriched the field of
computational invariant theory greatly \cite{BFGOWW19, DM20, MW19}. As
an immediate consequence, $\rit$ can also be solved in deterministic
polynomial time in the \emph{white-box} setting. Both problems admit a
randomized polynomial-time black-box algorithm due to Derksen and
Makam \cite{DM17}. Essentially, the result of \cite{DM17} shows that
to test whether a rational formula of size $s$ is zero or not (more
generally, whether a linear matrix of size $2s$ is invertible or not
over the free skew field), it suffices to evaluate the formula
(resp. the linear matrix) on random $2s\times 2s$ matrices.

Two central open problems in this area are to design faster
deterministic algorithms for the $\nsing$ problem and $\rit$ problem
in the black-box setting, raised in \cite{GGOW16, GGOW20}. The
algorithms in \cite{GGOW16} and \cite{IQS18} are inherently sequential
and seem unlikely to be helpful for designing a subexponential-time
black-box algorithm. Even for the $\rit$ problem (which could be
easier than the $\nsing$ problem), there is limited progress towards
designing an efficient deterministic black-box algorithm. A
deterministic quasipolynomial-time black-box algorithm is known for
identity testing of rational formulas of inversion height two
\cite{ACM22}. More recently, it is known that certain ABP (algebraic
branching program)-hardness of polynomial identities (PI) for matrix
algebras will yield a black-box subexponential-time derandomization of
$\rit$ in almost general setting \cite{ACGMR23}. However, the result
is conditional as such a hardness result is not proven. It is
interesting to note that in the literature of identity testing, the
$\nsing$ problem and the $\rit$ problem are rare examples of problems
with deterministic polynomial-time white-box algorithms but no known
deterministic subexponential-time black-box algorithms.

It is well-known \cite{GGOW16} that an efficient black-box algorithm (via a hitting
set construction) for $\nsing$ would generalize the celebrated quasi-$\NC$ algorithm for bipartite perfect
matching significantly \cite{FGT21}. This motivates the study of the parallel complexity of $\nsing$ and $\rit$. From the result of Derksen and Makam \cite{DM17}, one can observe that $\rit$ in the white-box setting can be solved in randomized $\NC$ which involve formula evaluation, and matrix operations (addition, multiplication, and inverse computation) \cite{Brent74, Csanky76, Berko84, HW15}.\footnote{Similarly $\nsing$ is also in randomized $\NC$ via the determinant computation \cite{Berko84, DM17}.} Designing a hitting set in quasi-$\NC$ for this problem would therefore yield a deterministic quasi-$\NC$ algorithm for this problem.
\subsection{Our Results}\label{ourresult}

In this paper, we focus on the RIT problem and improve the black-box complexity significantly by showing the following result.

\begin{theorem}\label{thm:main-theorem}
  For $n$-variate noncommutative rational formulas of size $s$ and inversion
  height $\theta$, we can construct a hitting set
  $\H_{n,s,\theta}\subseteq \M_{\ell_{\theta}}(\Q)^n$ of size $(ns)^{O(\theta^5 \log s)}$ in deterministic $ (ns)^{O(\theta^5 \log s)}$ time where $\ell_{\theta} = s^{O(\theta)}$.  \end{theorem}

Here $\M_{\ell_{\theta}}(\Q)$ denotes the $\ell_{\theta}$-dimensional matrix algebra over $\Q$.  An immediate corollary of \cref{thm:main-theorem} and \cref{fact:inv-height-bound} is the following.

\begin{corollary}[black-box $\rit$]\label{thm:main-theorem-bb}
In the black-box setting, $\rit$ can be solved in deterministic quasipolynomial time via an explicit hitting set construction. 
\end{corollary}

Note that even for noncommutative formulas i.e.\ when the inversion
height $\theta = 0$, the best known hitting set is of
quasipolynomial size and improving it to a polynomial-size hitting set
is a long standing open problem \cite{FS13}. In this light,
\cref{thm:main-theorem} is nearly the best result one can hope for,
apart from further improving the logarithmic factors on the exponent.


We further show that our hitting set construction can in fact be performed in quasi-$\NC$. In the white-box setting, we can evaluate a given rational formula on the hitting set points in parallel. This involves the evaluation of the formula in parallel, and supporting matrix addition, multiplication, and inverse computation. It is already observed that Brent's 
formula evaluation \cite{Brent74} can be adapted to the setting of noncommuatative rational formulas \cite{HW15}, and such matrix operations can be performed in $\NC$ \cite{Csanky76, Berko84}. Combining these results with the quasi-$\NC$ construction of the hitting set, we obtain the following corollary. 

\begin{corollary}[white-box $\rit$]\label{thm:main-theorem1}
In the white-box setting $\rit$ is in deterministic quasi-$\NC$. 
\end{corollary}

\subsection{Proof Idea}\label{sec:proofidea}

The main idea is to construct a hitting set for rational formulas of every inversion height inductively. Our goal is now to construct a hitting set for rational formulas of inversion height $\theta$ given a hitting set for formulas of height $\theta - 1$. To design a black-box RIT algorithm for rational formulas of size at most $s$, it suffices to construct a hitting set for rational formulas of inversion height at most $O(\log s)$, we can stop the induction at that stage.

As already mentioned, a noncommutative rational formula is nonzero in the free skew field if it evaluates to nonzero for some matrix substitution. However, the difficulty is that a rational formula may be undefined for a matrix substitution. For instance, this will happen if there is an inversion gate to which the input subformula evaluated to a singular matrix. Informally speaking, it is somewhat easier to maintain that nonzero subformulas evaluate to nonzero matrices, but it is much harder to ensure that they evaluate to \emph{non-singular} matrices. Therefore, a rational formula of inversion height $\theta$ may not even be defined on any matrix tuple in the hitting set of rational formulas of inversion height $\theta - 1$.

One way to handle this difficulty is to evaluate rational formulas on elements of a division algebra. Finite dimensional division algebras are associative algebras in which every nonzero elements are invertible.

This idea of embedding inside a division algebra proves to be very useful. We formally define the notion of hitting set for rational formulas (of any arbitrary inversion height) inside a division algebra.

\begin{definition}[Division algebra hitting set]
  For a class of $n$-variate rational formulas, a \emph{division algebra hitting set} is a hitting set for that class of formulas that is contained in $D^n$ for some finite-dimensional division algebra $D$.
\end{definition}

The advantage of a division algebra hitting set is that, whenever a rational formula evaluates to some nonzero value (at an $n$-tuple in the hitting set), the output is invertible.\footnote{Division algebras are used in \cite{IQS18} as part of the regularity lemma and their noncommutative rank algorithm. Roughly speaking, after each rank-increment step, if the matrix substitutions are $d$-dimensional these are replaced by $d$-dimensional division algebra elements. This is a \emph{rank rounding} step which ensures that the rank is always a multiple of $d$ after each increment. This is a white-box process as it crucially uses the structure of the linear matrix.  In contrast, our paper deals with the black-box setting where we do not have an explicit description of the rational formula (and the corresponding linear matrix).}

Therefore every $n$-variate rational formula of inversion height $\theta$ is defined on some $n$-tuple in any given division algebra hitting set for rational formulas of inversion height $\theta -1$. Can we build on this to efficiently construct a division algebra hitting set for rational formulas of inversion height $\theta$? In that case, we could inductively construct a division algebra hitting set for rational formulas of every inversion height. For the base case of the induction, we want to construct a division algebra hitting set for noncommutative formulas. An earlier work \cite{ACM22} shows this by essentially embedding the hitting set obtained by Forbes and Shpilka \cite{FS13} for noncommutative \emph{polynomials} computed by algebraic branching programs (ABPs) in a cyclic division algebra (see \cref{sec:cyclic} for the definition of a cyclic division algebra) of suitably small index i.e.\ the dimension of its matrix representation. The inductive construction of the division algebra hitting set is the main technical step we implement here using several conceptual and technical ideas.

At this point, we take a detour and examine the connection between the RIT and $\nsing$ problems. It is known that RIT is polynomial-time reducible to $\nsing$ \cite{HW15}. But do we need the full power of $\nsing$ to solve the RIT problem for rational formulas of height $\theta$ given a hitting set for formulas of height $\theta - 1$? Consider the following promise version of $\nsing$. The input is a linear matrix $T(x_1, \ldots, x_n)$ and a matrix tuple $(p_1, \ldots, p_n)\in D^n_{1}$ for a cyclic division algebra $D_{1}$. The promise is that there is a submatrix $T'$ of size $s-1$ (obtained by removing the $i^{th}$ row and $j^{th}$ column, for some $i,j\in [s]$) such that $T'(p_1, \ldots, p_n)$ is invertible.  It is easier to think such a tuple $(p_1, \ldots, p_n)$ as a \emph{witness}. 
The question is now to check the singularity of $T$ over the free skew field. We show that the construction of a hitting set for rational formulas of inversion height $\theta$ inductively reduces to this special case where the witness is some tuple in the hitting set for height $\theta - 1$.      
We then consider the shifted matrix $T(x_1 + p_1, \ldots, x_n + p_n)$.
Applying Gaussian elimination, we can convert the shifted matrix of form:
 \begin{equation}\label{eq:nice-form}
 U \cdot T(x_1 + p_1, \ldots, x_n + p_n) \cdot V =
\left[
\begin{array}{c|c}
I_{s-1} - L & A_j \\
\hline
B_i & C_{ij}
\end{array}
\right],
 \end{equation}
 where the entries of $L, A_j, B_i, C_{ij}$ are homogeneous $D_{1}$-linear forms. Here $B_i$ is a row vector and $A_j$ is a column vector. At a high level, this is conceptually similar to an idea useful for approximating commutative rank \cite{BBJP19}.
 
 It is not too difficult to prove that $T$ is invertible, if and only if, $C_{ij}-B_i(I_{s-1}-L)^{-1}A_j = C_{ij}-B_i(\sum_{k\geq 0}L^{k})A_j$ is a nonzero series.  By a standard result on noncommutative formal series \cite[Corollary 8.3]{Eilenberg74}, this is equivalent to saying that the truncated polynomial $C_{ij}-B_i(\sum_{k\leq {(s-1)\ell}}L^{k})A_j$ is nonzero where $\ell$ is the index of $D_{1}$. However, this series and the polynomial obtained will have division algebra elements interleaved with variables. Such a series (resp. polynomial) is called a generalized series (resp. generalized polynomial) and is studied extensively, for example, in the work of Vol\v{c}i\v{c} \cite{vol18}. We define
 similar notion of generalized ABPs and show that the truncated generalized polynomial of our interest is indeed computable by a polynomial-size generalized ABP. Finally, (up to a certain scaling by scalars) the upshot is that the division algebra hitting set construction for rational formulas is inductively reducible to the division algebra hitting set construction for such generalized ABPs.

We now consider such generalized ABPs where the coefficients lie inside a cyclic division algebra $D$ of index $\ell$, and call such ABPs as $D$-ABPs. Our goal is to construct a division algebra hitting set for such ABPs. To do so, a key conceptual idea is to introduce new noncommuting indeterminates for every variable and use the following mapping:
\[
x_i\mapsto \sum_{j,k = 1}^{\ell_1} C_{jk}\otimes y_{ijk},
\]
where the matrices $\{C_{jk}\}$ form a basis for the division algebra $D$. The idea is to overcome the problem of interleaving division algebra elements using the property of tensor products. This substitution reduces the problem to identity testing of a noncommutative ABP in the $\{y_{ijk}\}$ variables. Luckily, a division algebra hitting set construction for noncommutative ABPs is already known \cite{ACM22}. However, we need the hitting set to be inside a division algebra that contains $D$ as a subalgebra. A natural thought could be to take the tensor product of $D$ and the division algebra, say $D'$, that defines the division algebra hitting set for the noncommutative ABP in the $\{y_{ijk}\}$ variables.  However, in general, the tensor product of two division algebras is not a division algebra. At this point, we use a result \cite[Proposition, pg 292]{Pie82} that the tensor product of two cyclic division algebras of relatively prime indices $\ell_1$ and $\ell_2$, respectivly, is a cyclic division algebra of index $\ell_1\ell_2$. However, the division algebra hitting set construction \cite{ACM22} for noncommutative ABPs was only for division algebras whose index is only a power of two. Thus, in order to apply the above tensor product construction  \cite{Pie82} in several stages recursively, we need a division algebra hitting set construction for a division algebra whose index is a power of any \emph{arbitrary} prime $\p$.

We now informally describe how to construct such a division algebra hitting set for noncommutative formulas (more generally for noncommutative ABPs). For simplicity, suppose the prime is $\p$ and the ABP degree is $\p^d$. In \cite{FS13}, it is assumed that the degree of the ABP is $2^{d}$ and the construction has a recursive structure. In fact, itt is essentially a reduction to the hitting set construction for ROABPs (read-once oblivious algebraic branching programs) in commuting variables $u_1, u_2, \ldots, u_{2^d}$. The recursive step in their construction is by combining hitting sets (via hitting set generator $\mathcal{G}_{d-1}$) for two halves of degree $2^{d-1}$ \cite{FS13} with a rank preserving step of matrix products to obtain the generator $\mathcal{G}_d$ at the $d^{th}$ step. More precisely, $\G_d$ is a map from $\F^{d+1}\rightarrow \F^{2^{d}}$ that stretches the seed $(\alpha_1, \ldots, \alpha_{d+1})$ to a $2^d$ tuple for the read-once variables.

For our case, the high-level idea is to decompose the ABP of degree $\p^d$ in $\p$ consecutive parts each of length $\p^{d-1}$. We can adapt the rank preserving step for \emph{two matrix products} in \cite{FS13} to \emph{$\p$ many matrix products}. However, the main task is to ensure that the hitting set points lie inside a division algebra.    
For our purpose, we take a classical construction of cyclic division algebras \cite[Chapter 5]{Lam01}. The cyclic division algebra $D=(K/F,\sigma,z)$ is defined using a indeterminate $x$ as the $\ell$-dimensional vector space:
\[
D = K\oplus Kx\oplus \cdots \oplus Kx^{\ell-1},
\]
where the (noncommutative) multiplication for $D$ is defined by $x^\ell = z$ and $xb = \sigma(b)x$ for all $b\in K$.  Here $\sigma: K\rightarrow K$ is an automorphism of $K$ that fixes $F$, such that $\sigma$ generates the Galois group $\Gal(K/F)$. The fields we choose are $F=\Q(\omega_0,z)$ and $K=F(\omega)$, where $z$ is an indeterminate, $\omega_0$ is a root of unity whose order is relatively prime to $\p$, and $\omega$ is an $\ell^{th}$ primitive root of unity where $\ell$ is a suitable power of $\p$. The matrix
 representation of the division algebra element $xb$ for any $b\in K$ is of particular interest in this paper. It has the following circulant form: 
\[
\begin{bmatrix}
    0       & b & 0 & \cdots & 0 \\
     0       & 0 & \sigma(b) &\cdots  &  0 \\
      \vdots & \vdots &\ddots &\ddots  &  \vdots \\
      0       & 0 &\cdots                  & 0 &\sigma^{\ell-2}(b) \\
      z\sigma^{\ell-1}(b)       & 0 & \cdots & 0 & 0
\end{bmatrix}.
\]
Now, the broad idea to get the hitting set points inside a division algebra of index $\ell$ that is a power of $\p$ is as follows. Inductively assume that the construction follows the $\sigma$-automorphism (as in the matrix structure above) for each part of the generator output of length $\p^{d-1}$. Then we need to satisfy the $\sigma$-action at the $\p-1$ boundaries between the successive $\p^{d-1}$ length parts. In order to follow this inductive structure, we think of the division algebra construction as a tower of extension fields of $F$, with a higher-order root of unity at each stage.

Specifically, let $\omega_i =\omega^{\p^{a_i}}$ for $a_1>a_2>\cdots > a_d > a_{d+1} > 0$,
where $a_i$ are positive integers suitably chosen. Let $K_i=F(\omega_i)$
be the cyclic Galois extension for $1\le i\le d+1$ giving a tower
of extension fields
\[
F\subset F(\omega_1) \subset F(\omega_2)\subset \cdots \subset F(\omega_d) \subset F(\omega_{d+1}) \subset F(\omega).
\]

We require two properties of $\omega_i, 1\le i\le d+1$.
Firstly, for the hitting set generator $\G_i$ we will choose the root of
  unity as $\omega_i$ and the variable $\alpha_i$ will take values
  only in the set $W_i = \{\omega_0^{j'}\omega_i^j\mid 1\le j\le \p^{L-a_i} 0\le j'\le \ord{\omega_0}-1\}$.  We also require that for $1\le i\le d+1$ the map $\sigma^{\p^i}$ has $F(\omega_i)$ as its fixed field.

  The construction of matrix tuples in $D$ satisfying the above properties is the main technical step in \cref{hitting-set-p-divalg}. It turns out that choosing $\omega_0$ suitably plays an important role in the analysis of the final hitting set size (see Section~\ref{sec:main-result} for details).

We now conclude this section with a summary of the key steps involved in the hitting set construction. 
\subsubsection*{Informal summary.}
\begin{enumerate}
\item Division algebra hitting set construction for generalized ABPs defined over cyclic division algebras (\cref{sec:hit-gen-abps}).  
\begin{enumerate}
\item Given any prime $\p$, we build hitting set for noncommutative ABPs inside a cyclic division algebra whose index is a power of $\p$ (\cref{hitting-set-p-divalg}). 
\item We reduce the hitting set problem for generalized ABPs to that of noncommutative ABPs using the map $x_i \mapsto \sum C_{jk}\otimes y_{ijk}$ (the key idea in \cref{thm:gen-abp-hs}). 
\end{enumerate}

\item We construct a hitting set conditioned on a witness point for $\nsing$ problem (see Section~\ref{sec:nsing-witness} for details).  This builds on the construction of division algebra hitting set for generalized ABPs in Step 1 (\cref{thm:nsing-witness}). 

\item We construct the hitting set for rational formulas by induction on the inversion height $\theta$. The final hitting set construction and analysis are presented in \cref{sec:final-hs}. 
\end{enumerate}

\subsection{Related results}

The basic result underlying the hitting set construction in this paper is the Forbse-Shpilka hitting set construction~\cite{FS13}. At a high level, our approach builds on the framework introduced in \cite{ACM22, ACGMR23}. In \cite{ACM22}, the authors present a hitting set construction for rational formulas of inversion height two. The main ingredient in their construction is a division algebra hitting set construction for noncommutative formulas (more generally, for noncommutative ABPs). Additionally, they proposed an approach to constructing a hitting set by induction on inversion height as a possible approach to derandomize RIT in the black-box setting. Unfortunately they could not obtain a division algebra hitting set even for rational formulas of inversion height one. In \cite{ACGMR23}, the authors have shown a conditional result.
Assuming a conjecture on hardness of polynomial identities \cite{BW05} they
obtain a hitting set for every inversion height by induction. A limitation
of that construction is that it is based on an unproven conjecture, and
even so it only yields a quasipolynomial-size hitting set for rational formulas that is barely more than constant inversion height.

In this paper, we are able to overcome both the difficulties as we unconditionally build the quasipolynomial-size hitting set for \emph{all} polynomial-size rational formulas. 

As already mentioned, the results of \cite{GGOW20, IQS18, HH21} solve the more general $\nsing$ problem in order to solve the RIT problem in the white-box setting. 

In contrast, our hitting set construction crucially uses the inversion height of the input rational formula inductively. Furthermore, the hitting set construction for the $\nsing$ problem is known for the following special cases only: when the input matrix is a symbolic matrix (for which bipartite perfect matching is a special case) \cite{FGT21}, or more generally, when the input matrix consists of rank-1 coefficient matrices (for which linear matroid intersection is a special case) \cite{GT20}, and more recently, when the input matrix consists of rank-2 skew-symmetric coefficient matrices (fractional linear matroid matching is a special instance of it) \cite{GOR24}.\footnote{For the first two cases, invertibility over the (commutative) function field and invertibility over the (noncommutative) free skew field coincide.}

An exponential lower bound on the size of the rational formula computed as an entry of the inverse of a symbolic matrix is known \cite{HW15}. Therefore, our hitting set construction does not subsume these results. Similarly, it is quite unlikely to reduce the RIT problem (in the general setting) to any of these special cases of $\nsing$ problem. Thus it seems that these results are incomparable.

\subsection{Organization}
In \cref{sec:prelim}, we provide a background on algebraic complexity theory, cyclic division algebras, and noncommutative formal power series. 
hitting set for noncommutative ABPs over cyclic division algebras whose index is any prime power. In \cref{sec:hit-gen-abps} we give, in a self-contained presentation, the construction of a hitting set for generalized ABPs defined over cyclic division algebras. In \cref{sec:nsing-witness}, we obtain for the
$\nsing$ problem a hitting set conditioned on a witness point. The main result and analysis are in \cref{sec:main-result}. In \cref{sec:final-hs} we present the proof of our main result : a hitting set for arbitrary rational formulas (\cref{thm:main-theorem}). \cref{sec:quasinc} contains the proof of  \cref{thm:main-theorem1}. Finally, we raise a few questions for further research in \cref{sec:conclusion}.      

\section{Preliminaries}\label{sec:prelim}
\subsection{Notation}
Throughout the paper, we use $\F, F, K$ to denote fields, and ${\M_m(\F)}$ (resp. ${\M_m(F)}, {\M_m(K)}$) to denote $m$-dimensional matrix algebra over $\F$ (resp. over $F,K$).  Similarly, ${\M_m(\F)}^n$ (resp. ${\M_m(F)}^n, {\M_m(K)}^n$) denote the set of $n$-tuples over ${\M_m(\F)}$ (resp. ${\M_m(F)}, {\M_m(K)}$), respectively.  
$D$ is used to denote finite-dimensional division algebras. 
We use $\p$ to denote an arbitrary prime number.
Let $\ubar{x}$ denote the set of variables $\{x_1, \ldots, x_n\}$. Sometimes we use $\ubar{p} = (p_1, \ldots, p_n)$ and $\ubar{q} = (q_1, \ldots, q_n)$ to denote the matrix tuples in suitable 
matrix algebras where $n$ is clear from the context. The free noncommutative ring of polynomials over a field $\F$ is denoted by $\F\angle{\ubar{x}}$. 
For matrices $A$ and $B$, their usual tensor product is denoted by $A\otimes B$. For a polynomial $f$ and a monomial $m$, we use $[m]f$ to denote the coefficient of $m$ in $f$.   




\subsection{Algebraic Complexity Theory}\label{sec:alg-complexity}
\begin{definition}[Algebraic Branching Program]\label{abpdefn}
An \emph{algebraic branching program} (ABP) is a layered directed
acyclic graph. The vertex set is partitioned into layers
$0,1,\ldots,d$, with directed edges only between adjacent layers ($i$
to $i+1$). There is a \emph{source} vertex of in-degree $0$ in the layer
$0$, and one out-degree $0$ \emph{sink} vertex in layer $d$. Each edge
is labeled by an affine $\F$-linear form in variables, say, $x_1,x_2,\ldots,x_n$. The polynomial computed by
the ABP is the sum over all source-to-sink directed paths of the
ordered product of affine forms labeling the path edges. 
\end{definition}

The \emph{size} of the ABP is defined as the total number of nodes and the \emph{width} is the maximum number of nodes in a
layer, and the depth or length is the number of layers in the ABP.
An ABP can compute a commutative or a noncommutative polynomial, depending on whether the variables $x_1,x_2,\ldots,x_n$ occurring
in the $\F$-linear forms are commuting or noncommuting. ABPs of width $w$ can also be defined as an iterated matrix multiplication $ \ubar{u}^t\cdot M_1 M_2 \cdots M_{\ell} \cdot\ubar{v} $, where $\ubar{u}, \ubar{v}\in \F^n$ and each $M_i$ is of form $\sum_{i=1}^n A_ix_i$ for matrices $A_i\in \M_w(\F)$, assuming
without loss of generality that all matrices $M_j, 1\le j\le \ell$ are $w\times w$. Here, $\ubar{u}^{t}$ is the transpose of $\ubar{u}$.

We say a set $\mathcal{H}\subseteq \F^n$ is a hitting set for a (commutative) algebraic circuit class $\mathcal{C}$ if for every 
$n$-variate polynomial $f$ in $\mathcal{C}$, $f\not\equiv 0$ if and only if $f(\ubar{a}) \neq 0$ for some $\ubar{a}\in \mathcal{H}$.

A special class of ABPs in commuting variables are the \emph{read-once oblivious} ABPs (in short, ROABPs). In ROABPs a different variable is used for each layer, and the edge labels are univariate polynomials over that variable. 
For the class of ROABPs, Forbes and Shpilka~\cite{FS13} obtained the first quasipolynomial-time black-box algorithm by constructing a hitting set of quasipolynomial size.

\begin{theorem}{\rm\cite{FS13}}
For the class of polynomials computable by a width $r$, depth $d$, individual degree $< n$ ROABPs of known order, if $|\F| \geq (2dnr^3)^2$,
there is a $\poly(d, n, r)$-explicit hitting set of size at most $(2dn^2r^4)^{\lceil\log d + 1\rceil}$.
\end{theorem}

Indeed, they proved a more general result.

\begin{definition}[Hitting Set Generator]
A polynomial map $\mathcal{G} :\F^t \to \F^n $ is a generator for a circuit class $\mathcal{C}$ if for every $n$-variate polynomial $f$ in $\mathcal{C}$,
$f\equiv 0$ if and only if $f\circ \mathcal{G} \equiv 0$.
\end{definition}

\begin{theorem}~{\rm\cite[Construction~3.13, Lemma~3.21]{FS13}}~\label{theorem-FS13-ROABP-gen}
For the class of polynomials computable by a width $r$, depth $d$, individual degree $< n$ ROABPs of known order, one can construct a hitting set generator $\mathcal{G}: \F^{\lceil\log d + 1\rceil}\to \F^d$ of degree $dnr^4$ efficiently.
\end{theorem}


As a consequence, Forbes and Shpilka \cite{FS13}, obtain an
efficient construction of quasipolynomial-size hitting set for
noncommutative ABPs as well. Consider the class of noncommutative ABPs of
width $r$, and depth $d$ computing polynomials in $\F\angle{\ubar{x}}$. The
result of Forbes and Shpilka provide an explicit construction (in
quasipolynomial-time) of a set 
$\M_{d+1}(\F)$, such that for any ABP (with parameters $r$ and
$d$) computing a nonzero polynomial $f$, there always exists
$(p_1, \ldots, p_n)\in$$\mathcal{H}_{n, r, d}$, 
$f(\ubar{p})\neq 0$.

\begin{theorem}[Forbes and Shpilka \cite{FS13}]\label{forbesshpilka}
For all $n, r, d \in \mathds{N}$, if $|\F|\geq \poly(d,n,r)$, then there is a
hitting set $\mathcal{H}_{n, r, d} \subset \M_{d+1}(\F)$ for
noncommutative \text{ABP}s of parameters 
$|\mathcal{H}_{n, r,d}\mid \leq (rdn)^{O(\log d)}$ and there is a
deterministic algorithm to output the set $\mathcal{H}_{n, r, d}$ in
time $(rdn)^{O(\log d)}$.
\end{theorem}

\subsection{Cyclic Division Algebras}\label{sec:cyclic}
A division algebra $D$ is an associative algebra over a (commutative) field $\F$ such that all 
nonzero elements in $D$ are units (they have a multiplicative inverse). In this paper, we are interested in finite-dimensional division algebras. Specifically, we focus on cyclic division algebras and their construction \cite[Chapter 5]{Lam01}. Let $F=\Q(z)$, where $z$ is a commuting indeterminate. Let $\omega$ be an $\ell^{th}$ primitive root of unity. To be specific, let $\omega= e^{2\pi \iota/\ell}$. Let
$K=F(\omega)=\Q(\omega,z)$ be the cyclic Galois extension of $F$ obtained by
adjoining $\omega$. So, $[K:F]=\ell$ is the degree of the extension. The elements of $K$ are polynomials in $\omega$ (of
degree at most $\ell-1$) with coefficients from $F$.

Define $\sigma:K\to K$ by letting $\sigma(\omega)=\omega^k$ for some $k$
relatively prime to $\ell$ and stipulating that $\sigma(a)=a$ for all
$a\in F$. Then $\sigma$ is an automorphism of $K$ with $F$ as fixed
field and it generates the Galois group $\Gal(K/F)$.

The division algebra $D=(K/F,\sigma,z)$ is defined using a new
indeterminate $x$ as the $\ell$-dimensional vector space:
\[
D = K\oplus Kx\oplus \cdots \oplus Kx^{\ell-1},
\]
where the (noncommutative) multiplication for $D$ is defined by
$x^\ell = z$ and $xb = \sigma(b)x$ for all $b\in K$. 
The parameter $\ell$ is called the \emph{index} of $D$ \cite[Theorem 14.9]{Lam01}.  

The elements of $D$ has matrix representation 
in $K^{\ell\times \ell}$ from its action on the basis 
$\mathcal{X}=\{1,x,\ldots,x^{\ell-1}\}$. I.e., for $a\in D$ and $x^j\in\mathcal{X}$, the $j^{th}$ row of the matrix representation is obtained by writing $x^{j} a$ in the $\mathcal{X}$-basis. 

For example, the matrix representation $M(x)$ of $x$ is:

\[
        M(x)[i,j] = \begin{dcases}
                        1 & \text{ if } j=i+1, i\le \ell-1 \\
                        z & \text{ if } i=\ell, j=1\\
                        0 & \text{ otherwise.}
                    \end{dcases}
\]

$$
M(x)=\begin{bmatrix}
    0       & 1 & 0 & \cdots & 0 \\
     0       & 0 & 1 &\cdots  &  0 \\
      \vdots & \vdots &\ddots &\ddots  &  \vdots \\
      0       & 0 &\cdots                  & 0 &1 \\
      z       & 0 & \cdots & 0 & 0
\end{bmatrix}.
$$

For each $b\in K$ its matrix representation $M(b)$ is the diagonal matrix:

\[
        M(b)[i,j] = \begin{dcases}
                        b & \text{ if } i=j=1 \\
                        \sigma^{i-1}(b) & \text{ if } i=j, i\ge 2\\
                        0 & \text{ otherwise.}
                    \end{dcases}
\]

\[M(b) = 
\begin{bmatrix}
b & 0 & 0 & 0 & 0 & 0  \\
0 & \sigma(b) & 0 & 0 & 0 & 0 \\
0 & 0 & \sigma^2(b) & 0 & 0 & 0 \\
0 & 0 & 0 & \ddots & 0 & 0 \\
0 & 0 & 0 & 0 & \sigma^{\ell-2}(b) & 0 \\
0 & 0 & 0 & 0 & 0 & \sigma^{\ell-1}(b)
\end{bmatrix}
\]
        



\begin{proposition}\label{circ-in-D}
  For all $b\in K$, $M(bx) = M(b)\cdot M(x)$ 
\end{proposition}

Also, the matrix representation of $xb=\sigma(b)x$ is easy to see in the basis $\{1,x,\ldots,x^{\ell-1}\}$:
\[M(\sigma(b)x) = 
\begin{bmatrix}
0 & \sigma(b) & 0 & 0 & 0 & 0  \\
0 & 0 & \sigma^2(b) & 0 & 0 & 0 \\
0 & 0 & 0 & \sigma^3(b) & 0 & 0 \\
0 & 0 & 0 & 0 & \ddots & 0 \\
0 & 0 & 0 & 0 & 0 & \sigma^{\ell-1}(b) \\
\sigma^{\ell}(b)z & 0 & 0 & 0 & 0 & 0
\end{bmatrix}
\]

Define $C_{ij}= M(\omega^{j-1}) \cdot M(x^{i-1})$ for $1\leq i,j\leq \ell$. Observe that $\B=\{C_{ij}, i,j \in [\ell]\}$ is a $F$-generating set for the division algebra $D$. 

\begin{fact}\label{fact:div-alg}
  The $F$-linear span of $\B$ is the cyclic division algebra $D$ in the matrix algebra
  $\M_{\ell}(K)$.
\end{fact}

The following proposition is a standard fact. 
\begin{proposition}{\rm\cite[Section 14(14.13)]{Lam01}}\label{full-space}
The $K$-linear span of $\B$ is the entire matrix algebra 
$\M_{\ell}(K)$. 
\end{proposition}

The following theorem gives us a way of constructing new division algebras using tensor products. This construction
plays an important role in our main result. 

\begin{theorem}{\rm\cite[Proposition, Page 292]{Pie82}}\label{thm:cyclic-tensor}
Let $K, L$ be cyclic extensions of the field $F$ such that their extension degrees, $[K:F]$ and $[L:F]$, are relatively prime. 
Let $D_1=(K/F, \sigma_1, z)$, and $D_2=(L/F, \sigma_2, z)$ be the corresponding cyclic division algebras as defined above. 
Then their tensor product $D_1\otimes D_2$ is also a cyclic division algebra. 
\end{theorem}

\subsection{Noncommutative Rational Series}
Let $D$ be a division algebra and $P$ be a series over the noncommuting variables $x_1, x_2, \ldots, x_n$ defined as follows: 
\[
P={c} - {B}\left(\sum_{k\geq 0} L^k\right){A},
\]
where $c$ is a $D$-linear form (over $x_1, \ldots,x_n$), $B$ (resp. $A$) is a $1\times s$ (resp. $s\times 1$) dimensional vector, and $L$ is a $s\times s$ matrix. The entries of $B, L , A$ are $D$-linear forms over $x_1, \ldots, x_n$. Furthermore, the variables $x_i : 1\leq i\leq n$ commute with the elements in $D$.   
Define the truncated polynomial $\widetilde{P}$ as follows:
\begin{equation}\label{eqn:trunc-poly}
    \widetilde{P}=c - B\left(\sum_{k\leq s-1} L^k\right)A.
\end{equation}
The next statement shows that the infinite series $P\neq 0$ is equivalent in saying that $\widetilde{P}$ is nonzero. 
The proof of the fact is standard when $D$ is a (commutative) field \cite[Corollary 8.3, Page 145]{Eilenberg74}. For the case of division algebras, the proof can be found in \cite[Example 8.2, Page 23]{DK21}. However, we include a self-contained proof. 

\begin{fact}\label{lem:trucated}
The infinite series $P \neq 0$ if and only if its truncation $\widetilde{P}\neq 0$.   
\end{fact}

\begin{proof}
If $P=0$, then obviously $\widetilde{P}=0$, since the degrees in different homogeneous components do not match. 
Now, suppose $\widetilde{P}=0$. 
Notice that the terms in $c$ are linear forms and the degree of any term in $B\left(\sum_{k\geq 0} L^k\right)A$ is at least two. Hence, $c$ must be zero. 
Write the row and column vectors $B$ and $A$ as $B=\sum_{\ell} B_{\ell} x_{\ell}, A=\sum_{\ell} A_{\ell} x_{\ell}$. Similarly, write 
$L=\sum_{\ell} L_{\ell} x_{\ell}$. 

Suppose $B L^{s} A$ contributes a nonzero monomial (word) $w=x_{i_1}x_{i_2}\ldots x_{i_{s+2}}$. Clearly the coefficient of $w$ is $B_{i_1}L_{i_2}\ldots L_{i_{s+1}}A_{i_{s+2}}$. Consider the vectors  $v_1=B_{i_1}$, $v_2=B_{i_1} L_{i_2}$, \ldots, $v_{s+1}=B_{i_1} L_{i_2}\ldots L_{i_{s+1}}$ corresponding to the prefixes $w_1=x_{i_1}, w_2=x_{i_1}x_{i_2}, \ldots, w_{s+1}=x_{i_1}\ldots x_{i_{s+1}}$. These vectors $v_i, 1\le i\le s+1$ all lie in the (left) $D$-module $D^s$ which has rank $s$. As $D$ is a division algebra, these vectors cannot all be $D$-linearly independent. Hence, there are elements $\lambda_1,\ldots, \lambda_{s+1}$ in $D$, not all zero, such that the linear combination $\lambda_1 v_1 + \ldots +\lambda_{s+1} v_{s+1}=0$. However, $v_{s+1} A_{i_{s+2}}\neq 0$ by the assumption. Hence, there is at least one vector $v_{\ell} : 1\leq \ell \leq s$ such that $v_{\ell} A_{i_{s+2}}\neq 0$. This means that the coefficient of the word $w_{\ell}x_{i_{s+2}}$, which is of length at most $s+1$, is nonzero in $\widetilde{P}$, which is not possible by assumption. 

Now, with $k=s$ as the base case, we can inductively apply the above argument to show that 
$B L^k A$ is zero for each $k\ge s$.\end{proof}

\subsection{Generalized Formal Power Series} 

We now define the notion of generalized series first introduced by Vol\v{c}i\v{c}. For a detailed exposition, see~\cite{vol18}.  

A \emph{generalized word} or a \emph{generalized monomial} in $x_1,\ldots, x_n$ over the matrix algebra $\M_m(\F)$ allows the matrices to interleave between variables. That is to say, a generalized monomial is of the form: $a_0 x_{k_1}a_2\cdots a_{d-1}x_{k_d}a_{d}$, where $a_i\in \M_m(\F)$, and its degree is the number of variables $d$ occurring in it. A finite sum of generalized monomials is a \emph{generalized polynomial} in the ring $\M_m(\F)\angle{\ubar{x}}$. A \emph{generalized formal power series} over $\M_m(\F)$ is an infinite sum of generalized monomials such that the sum has finitely many generalized monomials of degree $d$ for any $d\in\mathds{N}$. The ring of
generalized series over $\M_m(\F)$ is denoted $\M_m(\F)\dangle{\ubar{x}}$.

A generalized series (resp. polynomial) $S$ over $\M_m(\F)$ admits the following canonical description. Let $E=\{e_{i,j}, 1\leq i,j\leq m\}$ be the set of elementary matrices. Express each coefficient matrix $a$ in $S$ in the $E$ basis by a $\F$-linear combination and then expand $S$. Naturally each monomial of degree-$d$ in the expansion looks like $e_{i_0,j_0} x_{k_1} e_{i_1,j_1} x_{k_2} \cdots e_{i_{d-1},j_{d-1}} x_{k_d} e_{i_d,j_d}$ where $e_{i_l,j_l}\in E$ and $x_{k_{l}}\in \ubar{x}$. We say the series $S$ (resp. polynomial) is identically zero if and only if it is zero under such expansion i.e. the coefficient associated with each generalized monomial is zero.   

The evaluation of a generalized series over $\M_m(\F)$ is defined on any $k'm\times k'm$ matrix algebra for some integer $k'\geq 1$ \cite{vol18}. To match the dimension of the coefficient matrices with the matrix substitution, we use an inclusion map $\iota: \M_m(\F)\to \M_{k'm}(\F)$, for example, $\iota$ can be defined as $\iota(a) = a\otimes I_{k'}$ or $\iota(a) = I_{k'}\otimes a$. Now, a generalized monomial  $a_0x_{k_1}a_1\cdots a_{d-1}x_{k_d}a_{d}$ over $\M_m(\F)$ on matrix substitution $(p_1,\ldots, p_n)\in \M_{k'm}(\F)^n$ evaluates to $$ \iota(a_0) p_{k_1} \iota(a_1)\cdots \iota(a_{d-1}) p_{k_d} \iota(a_d) $$under some inclusion map $\iota:\M_m(\F)\to \M_{k'm}(\F)$. All such inclusion maps are known to be compatible by the Skolem-Noether theorem~\cite[Theorem 3.1.2]{row80}. Therefore, if a series $S$ is zero with respect to some inclusion map $\iota: \M_m(\F)\to \M_{k'm}(\F)$, then it is zero w.r.t. any such inclusion map. Henceforth, we define the inclusion map as $\iota(a) = a\otimes I$ w.l.o.g.\ to evaluate a generalized series.

We naturally extend the definition of usual ABPs (\cref{abpdefn}) to the generalized ABPs. 
\begin{definition}[Generalized Algebraic Branching Program]\label{genabpdefn}
A \emph{generalized algebraic branching program} is a layered directed
acyclic graph. The vertex set is partitioned into layers
$0,1,\ldots,d$, with directed edges only between adjacent layers ($i$
to $i+1$). There is a \emph{source} vertex of in-degree $0$ in the layer
$0$, and one out-degree $0$ \emph{sink} vertex in layer $d$. Each edge
is labeled by a generalized linear form of $\sum_{i=1}^n a_i x_i b_i$ where $a_i, b_i \in\M_m(\F)$ for some integer $m$. As usual, \emph{width} is the maximum number of vertices in a layer. The generalized polynomial computed by
the ABP is the sum over all source-to-sink directed paths of the
ordered product of generalized linear forms labeling the path edges. 
\end{definition}

\begin{remark}
    It is clear from the definition above that such generalized ABPs with $d$ layers compute homogeneous generalized polynomials of degree $d$. 
\end{remark}

\section{Division Algebra Hitting Set for Generalized ABPs over Cyclic Division Algebras}\label{sec:hit-gen-abps}

In this section, we consider generalized ABPs where the coefficients are from a cyclic division algebra. We will construct a hitting set for such ABPs inside another cyclic division algebra. 

\begin{definition}[$D$-ABP]
Let $D=\left({K}/{F}, \sigma, z\right)$ be a cyclic division algebra of index $\ell$. We define a $D$-ABP as a generalized ABP $\mathcal{A}$ in $\{x_1, x_2, \ldots, x_n\}$ variables (as defined in \cref{genabpdefn}) where each edge is labeled by $\sum_{i=1}^n a_ix_ib_i : a_i, b_i\in D$. The ABP $\mathcal{A}$ computes a generalized polynomial over $D$.
\end{definition}

The main result of this section is a hitting set construction inside a cyclic division algebra for such division algebra ABPs. A key ingredient is the construction of a hitting set for noncommutative ABPs in a cyclic division algebra whose index is a power of any arbitrary prime $\p$. We first present this construction in the next subsection before addressing the general case.

\subsection{Division algebra hitting set for noncommutative ABPs}\label{appen-div-algebra-hitting-set}\label{subsec:divalg-hsg}

Fix a prime number $\p$. In particular, $\p$ is independent of the input ABP. In this section we show that the quasipolynomial-size hitting set construction for noncommutative ABPs by Forbes and Shpilka \cite{FS13} can be adapted to a more general setting where the hitting set points consist of matrices that lie in a finite-dimensional cyclic division algebra whose index is a power of $\p$. 
The construction here requires some more detail, especially as we will choose parameters keeping
in mind that the rational formula hitting set problem.


The Forbes -Shpilka construction gives a quasipolynomial size hitting set for ROABPs and hence obtains a similar size hitting set for noncommutative ABPs. In order to obtain a cyclic division algebra hitting set, we will take the Forbes-Shpilka hitting set construction for ROABPs and ensure additional conditions. We now present our ROABP hitting set construction. It is essentially based on \cite{FS13}. However, we present all the details, keeping it largely self-contained.

\subsubsection{Properties of the span of an ROABP}
Let $F$ be a characteristic zero field, which is a finite extension of
the function field $\Q(z)$ in indeterminate $z$. Let
$\{u_i\}_{0\leq i \leq \p-1}$ be $\p$ commuting indeterminates for
prime $\p$. The ring $\M_r(F[u_i])$ consists of $r\times r$ matrices
whose entries are univariate polynomials in $u_i$ over
$F$. Equivalently, any element $M\in\M_r(F[u_i])$ can be written as a
univariate polynomial $M=\sum_{j=0}^{d}M_ju_i^j$ in $u_i$ of degree
$\deg(M)=d$ with matrix coefficients $M_j\in\M_r(F)$ where $M_d\ne
0$. Let $\overline{F}$ denote the algebraic closure of $F$. The
following lemmas are useful generalizations of the results presented
in \cite[Subsection~3.1]{FS13}.

Let $\xi\in\overline{F}$ be a root of unity and let $K=F(\xi)$ denote
the field extension of $F$ by $\xi$. For matrices
$M_i\in\M_r(K), 1\le i\le t$ over a field $K$, let $\Span_K\{M_i\}$
denote the linear space $\{\sum_i\alpha_i M_i\mid \alpha_i\in K\}$.
The next two lemmas are a straightforward generalization of \cite[Lemma~3.2, Lemma~3.3]{FS13}, we skip their proofs.

\begin{lemma}\label{lem:span-product}{\rm\cite[Lemma~3.2]{FS13}}
  Let $K$ be any field.  For $1\le i \le t$ and $1\le \ell\le t'$
  let $M_{i\ell},M'_{i\ell} \in \M_r(K)$ be matrices such that
  $\Span_K\{M_{i\ell}\}_\ell = \Span_K\{M'_{i\ell}\}_\ell$ for each
  $i$. Then,
  \[
      \Span_K\left\{\prod_{i=0}^{\p-1} M_{i\ell}\right\}_\ell =
      \Span_K\left\{\prod_{i=0}^{\p-1} M'_{i\ell}\right\}_\ell.
    \]
\end{lemma}
  

\begin{lemma}\label{lem:FS-coeff-eval}
\[
\Span_{K}\left\{\left[\prod_{i=0}^{\p-1} u^{j_i}_i\right]\prod_{i=0}^{\p-1} M_i(u_i)\right\}_{\text{each~}j_i\in \{0,1, \ldots, n - 1\}} =
\Span_{K}\left\{\prod_{i=0}^{\p-1} M_i(u_i)\right\}_{\text{each~}u_i\in F}.
\]
\end{lemma}

\begin{lemma}\cite{GR08}\cite[Lemma~3.4]{FS13}\label{lem:gabison-raz}
  Let $\xi\in\overline{F}$ be a root of unity whose (finite) order
  $\ord(\xi) > \n$ and $K=F(\xi)$ be the field extension by $\xi$. Let
  $M\in \M_{\n,r}(F)$ For $\alpha \in K$, define
  $A_\alpha \in \M_{r,\n}(K)$ by $(A_\alpha)_{i,j} =
  (\xi^i\alpha)^j$. Then there are $< \n r$ values of $\alpha \in K$
  such that the first $r$ rows of $\rank(A_\alpha M) < r.$
\end{lemma}

\begin{lemma}\label{lem:GR-gen-FS}
  For each $i\in [\p]$, let $M_i(u_i) \in \M_r(F[u_i])$ be matrix
  polynomials of degree less than $\n$. Let $\xi\in\overline{F}$ be a
  root of unity such that $\ord(\xi) > \n^{\p}$ and let
  $K=F(\xi)$. Then for any $\alpha \in K$ and any $\mu \geq \n$,
\[
\Span_{K}\left\{\left[\prod_{i=0}^{\p-1} u^{j_i}_i\right]\prod_{i=0}^{\p-1} M_i(u_i)\right\}_{\text{each~}j\in \{0,1, \ldots, n - 1\}} \supseteq
\Span_{K}\left\{\prod_{i=0}^{\p-1} M_i((\xi^{\ell}\alpha)^{\mu^i})\right\}_{0\leq \ell\leq \ord(\xi)-1}.
\]
Moreover, for all but $\n^{p}r^2$ many values of $\alpha$ in $K$,
\[
\Span_{K}\left\{\left[\prod_{i=0}^{\p-1} u^{j_i}_i\right]\prod_{i=0}^{\p-1} M_i(u_i)\right\}_{\text{each~}j\in \{0,1, \ldots, n - 1\}} =
\Span_{K}\left\{\prod_{i=0}^{\p-1} M_i((\xi^{\ell}\alpha)^{\mu^i})\right\}_{0\leq \ell \leq r^2 - 1}.
\]
\end{lemma}

\begin{proof}
By definition we have

    \[
        \prod_{i=1}^{\p} M_i(u_i) = \sum_{j_i} \left(\left[\prod_{i=0}^{\p-1} u^{j_i}_i\right]
        \prod_{i=1}^{\p} M_i(u_i)\right )\cdot \prod_{i=0}^{\p-1} u^{j_i}_i.
    \]
Therefore, by substitution we have
\begin{equation}\label{eq:1}
  \prod_{i=1}^{\p} M_i((\xi^{\ell}\alpha)^{\mu^{i-1}}) = \sum_{j_i} \left(\left[\prod_{i=0}^{\p-1} u^{j_i}_i\right]\prod_{i=1}^{\p} M_i(u_i)\right)\cdot (\xi^{\ell}\alpha)^{j_0 + j_1\mu + \ldots + j_{\p-1}\mu^{\p-1}},
\end{equation}
which implies the first part of the lemma.

Now, we define a rectangular matrix $C\in \M_{\n^{\p},r^2}(F)$ as
follows. Each row of $C$ is indexed by a tuple
$(j_0,j_1,\ldots, j_{\p-1})\in \{0,1,\ldots, \n-1\}^{\p}$. For each such
tuple $(j_0,j_1,\ldots, j_{\p-1})$, treating the $r\times r$ matrix
$\left[\prod_{i=0}^{\p-1} u^{j_i}_i\right]\prod_{i=0}^{\p-1} M_i(u_i)$ as
an $r^2$-dimensional vector, we define it as the corresponding row
$C_{(j_0,j_1,\ldots, j_{\p-1})}$. By definition,
    \[
    \text{row-span}(C) = \Span\left\{\left[\prod_{i=0}^{\p-1} u^{j_i}_i\right]\prod_{i=0}^{\p-1} M_i(u_i)\right\}_{\text{each~}j \in \{0,1, \ldots, \n - 1\}}.
    \]
   
    Next, consider the rectangular matrix
    $A_{\alpha}\in \M_{r^2,\n^{\p}}(K)$ whose columns are indexed by
    tuples $(j_0,j_1,\ldots, j_{\p-1})\in \{0,1,\ldots, \n-1\}^{\p}$ with
    entries defined as
    \[
    (A_{\alpha})_{\ell, (j_0,j_1,\ldots, j_{\p-1})} = (\xi^{\ell}\alpha)^{j_0 + j_1\mu + \ldots + j_{\p-1}\mu^{\p-1}}.
    \]
    By \cref{lem:gabison-raz}, for all but $\n^{\p}r^2$ values of
    $\alpha$, we have $\rank(A_{\alpha}C) = \rank(C)$.  Multiplying
    the $\ell^{th}$ row of $A_{\alpha}$ with $C$ we get
    \[
    (A_{\alpha})_{\ell}C = \sum_{j_i} \left(\left[\prod_{i=0}^{\p-1} u^{j_i}_i\right]\prod_{i=0}^{\p-1} M_i(u_i)\right)\cdot (\xi^{\ell}\alpha)^{j_0 + j_1\mu + \ldots + j_{\p-1}\mu^{\p-1}} = \prod_{i=0}^{\p-1} M_i((\xi^{\ell}\alpha)^{\mu^{i}}).
    \]

    Therefore,
    \[\text{row-span}(A_{\alpha}C) = \Span\left\{M_0(\xi^{\ell}\alpha)M_1((\xi^{\ell}\alpha)^{\mu})\cdots M_{\p-1}((\xi^{\ell}\alpha)^{\mu^{\p-1}})\right\}_{0 \leq \ell\leq r^2 - 1\}}.
    \]

 As $\text{row-span}(C)$ contains $\text{row-span}(A_{\alpha}C)$, if
 $\rank(C) = \rank(A_{\alpha}C)$ then we have $\text{row-span}(C) =
 \text{row-span}(A_{\alpha}C)$. Therefore, for all but $\n^{\p}r^2$
 values of $\alpha$, $\text{row-span}(C) =
 \text{row-span}(A_{\alpha}C)$.
\end{proof}

%


We inductively assume that the ROABP is transformed
into an iterated matrix product of $\p^d$ matrices with the following
property. We can group this iterated matrix product into $\p$
consecutive sections, where the $i^{th}$ section is a product of
$\p^{d-1}$ consecutive matrices the entries of which are polynomial in
the variable $u_i, 0\le i\le \p-1$. Furthermore, this transformation
of the original ROABP is identity preserving, that is, the polynomial
computed by the original ROABP is nonzero if and only if a designated
entry of this matrix product is nonzero.





Then, broadly speaking, the next lemma gives a method to show that the
span of the full matrix product can be captured by span of the matrix
products over a \emph{single} variable. This will enable us to
transform the given iterated matrix product over variables
$\{u_i\}$ into an iterated matrix product over a single
variable. This lemma, based on Lagrangian interpolation, is a variant
of \cite[Lemma~3.7]{FS13}.


We recall the definition of Lagrange interpolation polynomials. Let
$\chi$ be a positive integer and
$S=\{\beta_0,\beta_1,\ldots,\beta_{\chi-1}\}$ be a set of $\chi$
distinct scalars. The Lagrange interpolation polynomials with respect
to $S$ is the set of $\chi$ many univariate polynomials
$q_{\ell,S}(v), 0\le \ell\le \chi-1$ of degree $\chi - 1$ in variable
$v$ defined as:
\[
q_{\ell, S}(v) = \prod_{k\neq \ell}\frac{v - \beta_k}{\beta_{\ell} -
  \beta_k} \quad\text{such that}\quad q_{\ell, S}(\beta_k)
= \begin{dcases} 1\quad \text{if~}\ell = k,\\ 0 \quad
  \text{otherwise}.
\end{dcases}
\]

\begin{lemma}\label{fs13-lemma-prime}
  Consider $\p$ many families of $r\times r$ matrices
  $\{M_{ij}\}_{0\leq j \leq p, 0\leq i \leq \p^{d-1}}$
  where for the $j^{th}$ family the entries are univariate polynomials
  over $F[u_j]$ of degree less than $n$.  Let
  $(f_0(u_j),f_1(u_j),\ldots,f_{\p^{d-1} - 1}(u_j))\in F[u_j]$ be polynomials of
  degree at most $m$ for each $0\leq j\leq \p-1$. Let $\omega$ be a primitive root of unity of
  order $\p^{L}$ for prime $\p>2$, where 
  $K=F(\omega)$. Let $\omega_0\in F$ be a root of unity of order
  $\Lambda={\mathscr{q}}^\tau > (\p^{d}nm)^{\p}$ where $\mathscr{q\neq \p}$ is a prime number. Let $1 \leq \gamma \leq \p^L$
  such that $\gamma$ is relatively prime to $\p^L$ and $\gamma \pmod \Lambda = 1$. Define polynomials
  in indeterminate $v$:
  \begin{eqnarray*}
    f'_{ij}(v) & = & \sum_{\ell = 0}^{r^2 -
                     1}\sum_{\ell'=0}^{\ord(\omega) - 1}
                     f_i(({\omega_0}^{\ell}{\omega}^{\ell'}
                     \alpha)^{\mu_{i,j}})q_{\ell\ell', S}(v)\\
   \end{eqnarray*}
  where $\mu_{i,j} = \mu^{j-1}\gamma^{i-1}$,
  and $q_{\ell\ell'}(v)$ denotes
  the Lagrange interpolation polynomials with respect to a set $S =
  \{\beta_{\ell\ell'}\}_{0\le \ell \le r^2-1, 0\le \ell'\le \ord(w)-1}$ where each $\beta_{\ell\ell'}\in F$.

 Then, for all but $(\p^{d-1}nm)^{\p}r^2$ many values of $\alpha$, 
 \[
 \Span_K\left\{\prod_{j=0}^{\p-1}\prod_{i=0}^{\p^{d-1} - 1} M_{ij}(f_i(u_j))\right\}_{\text{each~}u_j\in F} \subseteq 
 \Span_K\left\{\prod_{j=0}^{\p - 1}\prod_{i=0}^{\p^{d-1} - 1}M_{ij}(f'_i(v))\right\}_{v\in
   F}.
 \]
\end{lemma}  

\begin{proof}
    As before, all spans are $K$-linear spans. 
    For each $j$, let $R_j(u_j) = \prod_{i=0}^{\p^{d-1} - 1} M_{ij}(f_i(u_j))$.
    Note that $R_j(u_j)$ is a matrix of univariate polynomials in $u_j$ of degree less than $\p^{d-1} nm$. By definition, 
    \[
    \prod_{j=1}^{\p}\prod_{i=0}^{\p^{d-1} - 1} M_{ij}(f_i(u_j)) = \prod_{j=1}^{\p} R_j(u_j).
    \]
    \cref{lem:GR-gen-FS} and \cref{lem:FS-coeff-eval} imply that for
    $\mu > \p^dnm$, except for $<(\p^dnm)^pr^2$ many values of
    $\alpha$ in $K$,
\begin{equation}\label{eq:span-1}
\Span_{K}\left\{\prod_{j=0}^{\p - 1} R_j(u_j)\right\}_{\text{each~}u_j\in K} =
\Span_{K}\left\{\prod_{j=0}^{\p - 1} R_j((\omega_0\omega)^{\ell}\alpha)^{\mu^j})\right\}_{0\leq\ell\leq r^2 - 1}.
\end{equation}

\begin{claim}\label{FS-generalization-for-each-generator}

   \[
    \Span\left\{\prod_{j=0}^{\p - 1}\prod_{i=0}^{\p^{d-1} - 1} M_{ij}(f_i((\omega_0\omega)^{\ell}\alpha)^{\mu^j}))\right\}_{0\leq \ell \leq r^2 - 1} \subseteq \Span\left\{\prod_{j=0}^{\p-1}\prod_{i=0}^{\p^{d-1}-1} M_{ij}(f_i((\omega^{\ell}_0\omega^{\ell'}\alpha)^{\mu^j\gamma^{i}}))\right\}_{\substack{0 \leq \ell \leq r^2 - 1, \\ 0\leq \ell' \leq \ord(\omega) - 1}}.
    \] 
\end{claim}
\begin{proof}
  It suffices to show
  $\Span\left\{ M_{ij}(f_i((\omega^{\ell}\alpha)^{\mu^j}))\right\} =
  \Span\left\{
    M_{ij}(f_i((\omega^{\ell}_0\omega^{\ell'}\alpha)^{\mu^j\gamma^{i}}))\right\}$
  (we can then apply \cref{lem:span-product}). Let
  $\alpha = \omega^{j_1}_0\omega^{j_2}$. Therefore,
  $(\omega_0\omega)^{\ell}\alpha =
  \omega^{\ell+j_1}_0\omega^{\ell+j_2}$. Now, for each $i$ the map
  $a\mapsto a^{\gamma^i}$ for all
  $a\in \{\omega^\ell\mid 0\le \ell \le \ord(\omega)-1\}$ is a
  bijection as $\gamma$ is relatively prime to $\ord(\omega)$. Hence,
    \[
    ((\omega_0\omega)^{\ell}\alpha)^{\gamma^{-i}} = (\omega^{\ell+j_1}_0\omega^{\ell+j_2})^{\gamma^{-i}} = \omega^{\ell+j_1}_0(\omega^{\ell+j_2})^{\gamma^{-i}} = \omega^{\ell+j_1}_0 \omega^{\ell'+j_2} = \omega^{\ell}_0\omega^{\ell'}\alpha,
    \]
    for a unique $0 \leq \ell' \leq \ord(\omega)-1$, as $\gamma$ is
    relatively prime to the order of $\omega$.
\end{proof}

We now set $\mu_{i,j} = \mu^{j}\gamma^{i}$ in \cref{FS-generalization-for-each-generator} to get,
\begin{equation}\label{eq:span-2}
    \Span\left\{\prod_{j=0}^{\p-1}\prod_{i=0}^{\p^{d-1}-1} M_{ij}(f_i((\omega^{\ell}\alpha)^{\mu^j}))\right\}_{0\leq \ell \leq r^2 - 1} \subseteq \Span\left\{\prod_{j=0}^{\p-1}\prod_{i=0}^{\p^{d-1}-1} M_{j,i}(f_i((\omega^{\ell}_0\omega^{\ell'}\alpha)^{\mu_{i,j}}))\right\}_{\substack{0 \leq \ell \leq r^2 - 1, \\ 0\leq \ell' \leq \ord(\omega) - 1}}.
\end{equation}
    
    For each $j$, let $T_j(v) = \prod_{i=0}^{\p^{d-1}-1} M_{ij}(f'_{i + jp^{d-1}}(v))$. By the definition of the Lagrange interpolation polynomials, letting $q_\ell(\beta_k)=\delta_{\ell k}$ where each $\beta_k$ is distinct,
    we have
    \begin{equation}\label{eq:span-3}
    T_j(\beta_{\ell\ell'}) = \prod_{i=0}^{\p^{d-1}-1} M_{ij}(f_i((\omega^{\ell}_0\omega^{\ell'}\alpha)^{\mu_{i,j}})),
    \end{equation}
    where $(\omega_0\omega)^\ell\alpha = (\omega^{\ell}_0\omega^{\ell'}\alpha)^{\gamma^i}$.
    Combining \cref{eq:span-1}, \cref{eq:span-2}, and \cref{eq:span-3}, we obtain:
    \[
    \Span\left\{\prod_{j=0}^{\p-1} R_j(u_j)\right\}_{\text{each~}u_j \in F} \subseteq \Span{\left\{\prod_{j=0}^{\p-1} T_j(v)\right\}_{v\in F}}.\qedhere
    \]
\end{proof}

\begin{remark}
    \cref{fs13-lemma-prime} can be seen as a generalization of \cite[Lemma~3.7]{FS13}. However, here we use two different roots of unity $\omega_0$ and $\omega$ unlike \cite{FS13}. The order of $\omega_0$ is sufficiently large which allows us to use \cref{lem:GR-gen-FS} in order to prove the span containment, similar to \cite{FS13}. Whereas $\omega$ plays an important role to embed the hitting set points for the noncommutaitve ABPs inside a cyclic division algebra as shown in \cref{subsubsec:divalg-hsg}.
\end{remark}

\subsubsection{Construction of the ROABP hitting set generator}\label{sec:ROABP-hsg}

We will now proceed to describe the construction of a hitting set generator for ROABPs. This in turn will yield a
quasipolynomial size hitting set for noncommutative ABPs,
where the matrix substitutions will be from a cyclic division
algebra. Our construction and choice of parameters here is tailored
towards the main aim of the paper, which is obtaining a
quasipolynomial size hitting set for noncommutative rational formulas.
That we shall describe in subsequent sections.

Fix a prime $\p>2$ and integers $n, r, d$. Let $\omega=e^{\frac{2\pi
    \iota}{\p^L}}$, a root of unity of order $\p^L$ for some even
number $L > d$. Let $F=\Q(z, \omega_0)$ where $\omega_0$ is a root of
unity of order $\Lambda={\mathscr{q}}^\tau > (\p^dmn)^\p\cdot r^2$ where $\mathscr{q}\neq \p$ is a prime number and $K=F(\omega)$ is
its (finite) extension by $\omega$. Consider the $K$-automorphism
$\sigma:a\mapsto a^{1 + \Lambda \p^{\kappa}}$ where the positive integer
$\kappa = L/2$. It is easy to check that
$\sigma(\omega_0)=\omega_0$. As $\sigma(\omega)=\omega^{1+\Lambda
  \p^\kappa}$ and $1+\Lambda\p^\kappa$ is relatively prime to $\p^L$
it follows that $\sigma$ generates all the automorphisms of $K$ that
fix $F$.

Let $\omega_i =\omega^{\p^{\kappa - i}}$ for each $1\leq i \leq
d$. We denote by $K_i$ the cyclic Galois extension $K_i=F(\omega_i)$
of $F$ by $\omega_i$, for $1\le i\le d$.  This gives a tower of
field extensions
\[
F\subset F(\omega_1) \subset F(\omega_2)\subset \cdots \subset F(\omega_d)\subset F(\omega)=K.
\]

\begin{observation}
    For each $1\leq i \leq d$, $\sigma^{\p^i}$ fixes $F(\omega_i)$.
\end{observation}
\begin{proof}
    Define $a_i = \kappa - i$. As $\sigma(\omega)=\omega^{\Lambda \p^{\kappa}+1}$,
we have $\sigma(\omega_i) = \omega^{\p^{a_i}(\Lambda \p^{\kappa}+1)}$. Therefore,
\[
\sigma^{\p^i}(\omega_i) = \omega^{\p^{a_i}(\Lambda \p^{\kappa}+1)^{\p^i}}.
\]

Now, $(\Lambda \p^{\kappa}+1)^{\p^i}=\sum_{j=0}^{\p^i}{\binom{\p^i}{j}} \Lambda^{j} \p^{\kappa j}$. As
$\kappa=L/2$, we have $\omega^{\p^{\kappa j}}=1$ for $j\ge 2$. Therefore,
\[
  \sigma^{\p^i}(\omega_i) = \omega^{\p^{a_i}(\Lambda \p^{i+\kappa}+1)} =
  \omega_i\cdot \omega^{\Lambda \p^{a_i+i+\kappa}}.
\]

The choice of $a_i$, $a_i+i+\kappa = L$ for $1\leq i\le d$, implies
that $\omega^{\Lambda \p^{a_i+i+\kappa}}=1$ ensuring that
$\sigma^{\p^i}$ fixes $\omega_i$. \end{proof}

For each $1 \leq i \leq d$, define the set \[W_i = \{\omega^{j_1}_0\omega^{j_2}_i\mid 1\leq j_1\leq \Lambda,~ 1\le j_2\le \p^{\kappa + i}\}.\]

\noindent\textbf{Hitting Set Generator.} We now proceed to define the
hitting set generator as a mapping $\G_d:K^{d+1}\to K^{\p^d}$. The map
$\G_d$ will be recursively defined in terms of $\G_{d-1}$ which is for
matrix products of length $\p^{d-1}$. More generally, we will define
$\G_i: K^{i+1}\to K^{\p^i}$ in terms of $\G_{i-1}$, $1\le i\le d$.

For $0\le k\le \p^i-1$, we will denote the $k^{th}$ coordinate of the
output of $\G_i$ as the function $\G_{i,k}:K^{i+1}\to K$.
Lemma~\ref{fs13-lemma-prime} will play a role in the construction.
Specifically, the choice of the interpolating scalar set
$S_i\subseteq F$ used in the lemma and the effect of the automorphism
$\sigma$ on the Lagrange interpolation polynomials will be
important. In the definition of $\G_i$, we will choose a subset
$S_i\subset \mathcal{S}=\{\omega_0^\ell\mid 0\le \ell \le \Lambda-1\}$
of size $|S_i|=r^2\cdot \ord(\omega_i)=r^2\cdot \p^{\kappa +i}$. By
choice $\Lambda=\mathscr{q}^\tau> r^2\cdot\ord(\omega_i)$ for $1\le i\le d$.
Notice that $S_i\subset W_i$ and each element of $S_i$ is fixed
by $\sigma$.


For $0 \leq k \le \p^i-1$, we write $k = j\p^{i-1} + k'$ for some $0
\leq j \le \p-1$ and $0 \leq k' \le \p^{i-1}$. Define
$\G_{0,k}(\alpha_1)=\alpha_1$ and for $i>0$ define the mapping
$\G_{i,k}$ recursively as:
\begin{equation}\label{eq:generator-definition}
    \G_{i,k}(\alpha_1, \ldots, \alpha_i, \alpha_{i+1})
= 
\sum_{\ell'=0}^{\ord(\omega_i)-1}\sum_{\ell = 0}^{r^2 - 1} \G_{i-1,k'}(\alpha_1, \ldots, \alpha_{i-1}, (\omega^\ell_0\omega^{\ell'}_i\alpha_{i})^{\mu_{i}^j\gamma^{k'}}) q_{\ell\ell',S_i}(\alpha_{i+1}),
\end{equation}

where $\mu_i=1+\Lambda\p^{i-1+\kappa}$ and $\gamma = 1 + \Lambda \p^\kappa$.


\begin{remark}\label{rem:fs13}
 It is useful to note that for all $\alpha_i\in W_i$, from the choice of $\mu$ and $\gamma$, we have $\alpha^{\gamma^k}_i=\alpha^{\mu^j\gamma^{k'}}_i$. Hence in  \cref{fs13-lemma-prime}, we can equivalently write $(\omega^\ell_0\omega^{\ell'}_i\alpha_{i})^{\gamma^{k}}$ instead of $(\omega^\ell_0\omega^{\ell'}_i\alpha_{i})^{\mu_{i}^j\gamma^{k'}}$.
 \end{remark}

 \begin{remark}\label{rem:degree}
  We briefly explain the parameters $\mu_{i}^j$ and $\gamma^k$. The
  parameter $\mu_i^j$ is from setting $\mu$ in \cref{fs13-lemma-prime}
  as $\mu_i=1+\Lambda\p^{i-1+\kappa}$. Then
  $\mu_i^j=(1+\Lambda\p^{i-1+\kappa})^j$. Notice that if
  $\alpha_i\in W_i$ then
  $\ord(\alpha_i)\le |W_i|=\Lambda\cdot \ord(\omega_i)$ and, as
  observed in Remark~\ref{rem:fs13},
  $\alpha_i^{\mu_i^j\gamma^{k'}}=\alpha_i^{\gamma^k}$. Letting
  $e_{ik}=\gamma^k(\mod~\Lambda \ord(\omega_i))$ we can
  write  $\alpha_i^{\gamma^k}=\alpha_i^{e_{ik}}$ for each $i$,
  where $e_{ik}\le \Lambda \ord(\omega_i)-1$.   This implies that the degree of $\alpha_i$ for $i\le d$ in the polynomial
  $\G_d$ can be bounded by $\Lambda\ord(\omega_i)-1$.


\end{remark}  

\begin{lemma}\label{lem:hsg-span}
  For a positive integer $d$ let $M_k\in F[u_k]^{r\times r}$ for
  $0 \leq k \leq \p^d-1$, be matrices whose entries are polynomials of
  degree at most $n$. Then the following containment holds for the
  mapping $\G_d:K^{d+1}\to K^{\p^d}$ defined in
  \cref{eq:generator-definition}:
\[
\Span_K\left\{\prod_{k = 0}^{\p^d-1} M_k(u_k)\right\}_{\text{each~}u_k\in F} \subseteq
\Span_K\left\{\prod_{k=0}^{\p^d-1} M_k(\G_{d,k}(\alpha_1, \ldots, \alpha_d,
  \alpha_{d+1}))\right\}_{\substack{\alpha_{d+1}\in \mathcal{S}, \\ \text{each~}\alpha_i\in W_i, 1\le i\le d}}.
\]
\end{lemma}

\begin{proof}
  We will prove this by a suitable induction using
  \cref{fs13-lemma-prime}.  In order to set up the induction we
  observe that the $\p^d$ product of the matrices $M_k(u_k)$ (in
  $\p^d$ distinct commuting variables $u_k$) can be grouped into
  $\p$-products of matrices at a time forming a full $\p$-ary tree of
  depth $d+1$. Counting from the bottom layer numbered as layer $0$,
  at the $i^{th}$ layer of this $\p$-ary tree of matrix products we
  have $\p^{d-i}$ nodes, where each node represents a $\p^i$ matrix
  product. For $0\le k\le \p^d-1$, we can uniquely express
  $k=j\p^i+k'$ for $0\le k'\le p^i-1$ and $0\le j\le
  \p^{d-i}-1$. Denoting, for convenience, the matrix $M_{k}$ as
  $M_{jk'}$ we can write
\[
\prod_{k=0}^{\p^d-1}M_k(u_k)=\prod_{j=0}^{\p^{d-i}-1}\prod_{k'=0}^{\p^i-1}M_{jk'}(u_{k}).
\]

Now, we can state the claim which we prove by induction and obtain the
lemma. We want to show for $0\le i\le d$ that
\begin{equation}\label{eq:hsg}
\Span_K\left\{\prod_{k = 0}^{\p^d - 1} M_k(u_k)\right\}_{\text{each~}u_k\in F}
\subseteq \Span_K\left\{\prod_{j=0}^{\p^{d-i}-1}\prod_{k'=0}^{\p^i-1}
M_{jk'}(\G_{i,k}(\alpha_1, \ldots, \alpha_i,
\hat{u}_j)))\right\}_{\substack{\alpha_\ell\in W_\ell, ~1\le \ell \le i,\\
    ~\hat{u}_j\in \mathcal{S}}}.
\end{equation}
\noindent\textbf{Base Case: $i = 0$.}  For the base case $i=0$ we note
that the right hand side coincides with the left because for $i=0$ we
have $j=k$ and $\G_{0,k}(\hat{u}_j)=\hat{u}_k$
for $\hat{u}_k\in F$. Then as the entries of $M_k$ are polynomials of
degree at most $n$ in $u_k$ and $|\mathcal{S}|=\Lambda$ is
sufficiently larger than $n$ by choice, we can apply
\cref{lem:GR-gen-FS} (choosing $\xi=\omega_0$ and
$\alpha\in \mathcal{S}$ in that lemma)
\[
\Span_K\left\{\prod_{k = 0}^{\p^d-1} M_k(u_k)\right\}_{\text{each~}u_k\in F}
\subseteq \Span_K\left\{\prod_{k = 0}^{\p^d-1} M_k(u_k)\right\}_{\text{each~}u_k\in \mathcal{S}}.
\]



\noindent\textbf{Induction Step} Now suppose the containment of
Equation~\ref{eq:hsg} holds for some $i<d$. Our aim is to prove it for
$i+1$. From the induction hypothesis, it suffices to show the containment of the right hand side of
Equation~\ref{eq:hsg} in the span
\[
\Span_K\left\{\prod_{j'=0}^{\p^{d-i-1} - 1}\prod_{k''=0}^{\p^{i+1}-1}
M_{j'k''}(\G_{i+1,k}(\alpha_1, \ldots, \alpha_{i+1},
\hat{u}_j)))\right\}_{\substack{\alpha_\ell\in W_\ell,~1\le \ell \le i,\\
    \hat{u}_j\in F}}.
\]
Here, as we did for layer $i$, we are writing $k=j'\p^{i+1}+k''$ where
$0\le k''\le p^{i+1}-1$ and $0\le j'\le \p^{d-i-1}-1$. Also, we are
denoting the matrix $M_{k}$ as $M_{j'k''}$. We are going to prove this
containment by applying \cref{fs13-lemma-prime}. It is helpful to
write the RHS of Equation~\ref{eq:hsg} as the three-fold product
\[
\Span_K\left\{\prod_{j'=0}^{\p^{d-i-1} - 1}\prod_{j_1=0}^{\p-1}\prod_{k'=0}^{\p^i-1}
M_{jk'}(\G_{i,k}(\alpha_1, \ldots, \alpha_i,
\hat{u}_j)))\right\}_{\substack{\alpha_\ell\in W_\ell, ~1\le \ell \le i,\\
    \hat{u}_j\in F}},
\]
where the index $j$ used in RHS of Equation~\ref{eq:hsg} is
essentially split into $j'$ and $j_1$ where $j=j'\p^{d-i - 1}+j_1$.
Likewise, note that we can also write the index $k''$ as
$k''=j_1\p^i+k'$ for $0\le k''\le p^{i+1}-1$. Then, by the short-hand
we are using, the matrix $M_k$ is the same as $M_{jk'}$ and
$M_{j'k''}$.
 
 By \cref{lem:span-product} it suffices to show for each
$0\le j'\le \p^{d-i-1}$ and each fixed value of
$(\alpha_1,\alpha_2,\ldots,\alpha_{i})\in W_1\times W_2\times \cdots
\times W_{i}$:
$\Span_K\left\{\prod_{j_1=0}^{\p-1}\prod_{k' = 0}^{\p^i-1}
  M_{j'k''}(\G_{i,k}(\alpha_1, \ldots, \alpha_{i},
  \hat{u}_{j_1}))\right\}_{\hat{u}_{j_1}\in F}$ is contained in
$\Span_K\left\{\prod_{k'' = 0}^{\p^{i+1}-1}
  M_{j'k''}(\G_{i+1,k}(\alpha_1, \ldots, \alpha_{i+1},
  v))\right\}_{\alpha_{i+1}\in W_{i+1}, v\in
  \mathcal{S}}$.

Each matrix entry in the first span above is a univariate polynomial
over $\hat{u}_{j_1}$ only. Therefore, by \cref{lem:span-product} and
\cref{fs13-lemma-prime} we obtain,
\begin{align*}
&\Span_K\left\{\prod_{j_1=0}^{\p-1}\prod_{k' = 0}^{\p^i-1}
  M_{j'k''}(\G_{i,k}(\alpha_1, \ldots, \alpha_i,
  \hat{u}_{j_1}))\right\}_{\hat{u}_{j_1}\in F} \\&\subseteq \Span_K\left\{\prod_{j_1=0}^{\p-1}\prod_{k' = 0}^{\p^i-1}
  M_{j'k''}\left(\left(\sum_{\ell'=0}^{\ord(\omega_{i+1})-1}\sum_{\ell =0}^{r^2 - 1}\G_{i,k}(\alpha_1, \ldots,
  \alpha_{i}, ((\omega_0^{\ell}
  \omega_{i+1}^{\ell'}\alpha_{i+1})^{\mu_i^{j_1}\gamma^{k'}})\right)q_{\ell\ell',S_i}(v)\right)\right\}_{\alpha_{i+1\in W_{i+1},}v\in \mathcal{S}}\\
        &\subseteq
        \Span_K\left\{\prod_{k'' = 0}^{\p^{i+1}-1}
  M_{j'k''}(\G_{i+1,k}(\alpha_1, \ldots, \alpha_{i+1},
  v))\right\}_{\alpha_{i+1}\in W_{i+1} ,v\in \mathcal{S}}. 
\end{align*}
Importantly, by \cref{fs13-lemma-prime} the number of bad choices for
$\alpha_{i+1}\in W_{i+1}$ is only a small fraction of $|W_{i+1}|$ by
our choice of $\Lambda$.

\end{proof}

\begin{lemma}
  The mapping $\G_d:K^{d+1}\to K^{\p^d}$ as defined in
  \cref{eq:generator-definition} has the property that for
  $(\alpha_1,\alpha_2,\ldots,\alpha_{d+1})\in W_1\times
  W_2\times\cdots \times W_d\times \mathcal{S}$ and for any
  $1\leq k < \p^d$
\[
\sigma(\G_{d,k}(\alpha_1, \ldots, \alpha_d, \alpha_{d+1})) = \G_{d,k+1}(\alpha_1, \ldots, \alpha_d, \alpha_{d+1}).
\]
\end{lemma}

We need to show for $\alpha_i \in W_i$, $1\leq i \leq d$ and
$\alpha_{d+1}\in \mathcal{S}$, for any $1\leq k < \p^d$
\[
\sigma(\G_{d,k}(\alpha_1, \ldots, \alpha_d, \alpha_{d+1})) = \G_{d,k+1}(\alpha_1, \ldots, \alpha_d, \alpha_{d+1}).
\]

The proof is by induction with some tedious details.
Recall that $\sigma(\omega) = \omega^{\gamma}$. For readability we
shall use $\ubar{\alpha}$ for the $(d-2)$-tuple
$(\alpha_1,\alpha_2,\ldots,\alpha_{d-2})$. Inductively, suppose for
any $\alpha_i \in W_i$ for $1\leq i \leq d-1$ and
$\alpha_{d}\in \mathcal{S}$, and for $1\leq k' < \p^{d-1}$ we have
$\sigma(\G_{d-1,k}(\ubar{\alpha}, \alpha_{d-1}, \alpha_{d})) =
\G_{d-1,k+1}(\ubar{\alpha}, \alpha_{d-1}, \alpha_{d})$. Then by
definition of $\G_{d-1}$ (Equation~\ref{eq:generator-definition}),
writing $k=j\p^{d-1}+k'$ we have
\begin{align*}
  \sigma\left(\sum_{\ell,\ell'} \G_{d-2,k}(\ubar{\alpha},
  (\omega_0^{\ell}\omega^{\ell'}_{d-1}\alpha_{d-1})^{\mu_{d-1}^j\gamma^{k'}})
  q_{\ell\ell',S_{d-2}}(\alpha_d)\right) &= \sum_{\ell,\ell'}
                              \G_{d-2,k+1}(\ubar{\alpha},
                              (\omega_0^{\ell}\omega^{\ell'}_{d-1}\alpha_{d-1})^{\mu_{d-1}^j\gamma^{k'+1}}) q_{\ell\ell',S_{d-2}}(\alpha_d).
\end{align*}
For $\alpha_d\in \mathcal{S}$ we have $\sigma(\alpha_d) = \alpha_d$. Hence,
we have for each $\ell$ and $\ell'$,
\[
\sigma(\G_{d-2,k}(\alpha_1, \ldots, (\omega_{d-1}^{\ell}\alpha_{d-1})^{\gamma^k}) = \G_{d-2,k+1}(\alpha_1, \ldots, (\omega_{d-1}^{\ell}\alpha_{d-1})^{\gamma^{k+1}}).
\]

At this point, it is useful to introduce some notation for ease of
reading. We will assume $(\alpha_1,\alpha_2,\ldots,\alpha_{d-2})$ is
a fixed $(d-2)$-tuple. Let $f'_k(\alpha_{d-1},\alpha_d,\alpha_{d+1})$,
$g_k(\alpha_{d-1},\alpha_d)$ and $h_k(\alpha_{d-1})$ denote the
maps $\G_{d,k}(\ubar{\alpha}, \alpha_{d-1},\alpha_d,\alpha_{d+1})$,
$\G_{d-1,k}((\ubar{\alpha}, \alpha_{d-1},\alpha_d)$, and
$\G_{d-2,k}((\ubar{\alpha}, \alpha_{d-1})$ respectively. 
We now define $G_{d,k}$ in terms of $\G_{d-2,k}$ using the above
notation using Remark~\ref{rem:fs13} that
$a^{\gamma^k}=a^{\mu_i^j\gamma^{k'}}$ for $a\in W_i$ where
$k=j\p^{i-1}+k'$ and $k'=j'\p^{d-2}+k''$.

\begin{align*}
  \G_{d,k}(\alpha_1, \ldots, \alpha_d, \alpha_{d+1})
  &=  f'_k(\alpha_{d-1}, \alpha_d, \alpha_{d+1})\\
  &= \sum_{\ell,\ell'} g_{k'}(\alpha_{d-1},
    (\omega_0^\ell\omega^{\ell'}_d\alpha_{d})^{\gamma^{k}}) q_{\ell\ell',S_d}(\alpha_{d+1})\\
  &= \sum_{\ell,\ell'}
    \left(\sum_{\ell_1,\ell_2}h_{k''}((\omega_0^{\ell_1}\omega^{\ell_2}_{d-1}\alpha_{d-1})^{\gamma^{k'}}) q_{\ell_1\ell_2,S_{d-1}}((\omega_0^{\ell_1}\omega^{\ell_2}_d\alpha_{d})^{\gamma^k})\right)q_{\ell\ell',S_d}(\alpha_{d+1}).
\end{align*}


\begin{align*}
\text{Therefore,~~}&  \sigma(\G_{d,k}(\alpha_1, \ldots, \alpha_d, \alpha_{d+1}))
   = \sigma(f'_k(\alpha_{d-1}, \alpha_d, \alpha_{d+1}))\\
    &= \sum_{\ell,\ell'}
    \left(\sum_{\ell_1,\ell_2}\sigma(h_{k''}((\omega_0^{\ell_1}\omega^{\ell_2}_{d-1}\alpha_{d-1})^{\gamma^{k'}})
    q_{\ell_1\ell_2,S_{d-1}}(\sigma(\omega_0^\ell\omega^{\ell'}_d\alpha_{d}^{\gamma^k}))\right)
      q_{\ell\ell',S_d}(\sigma(\alpha_{d+1}))\\
  &= \sum_{\ell,\ell'}
    \left(\sum_{\ell_1,\ell_2}
    h_{k_1}((\omega_0^{\ell_1}\omega^{\ell_2}_{d-1}\alpha_{d-1})^{\gamma^{k'+1}})
    q_{\ell_1\ell_2,S_{d-1}}((\omega_0^\ell\omega^{\ell'}_d\alpha_{d})^{\gamma^{k+1}}))\right)
    q_{\ell\ell',S_d}(\alpha_{d+1})\\
    &= \sum_{\ell,\ell'}
      \left(\sum_{\ell_1,\ell_2}\G_{d-2,k_1}(\ubar{\alpha},(\omega_0^{\ell_1}\omega^{\ell_2}_{d-1}\alpha_{d-1})^{\mu_{d-1}^{j'}\gamma^{k''+1}})
      q_{\ell_1\ell_2,S_{d-1}}((\omega_0^\ell\omega^{\ell'}_d\alpha_{d})^{\mu_d^j\gamma^{k'+1}})\right)q_{\ell\ell',S_d}(\alpha_{d+1})\\
  &= \G_{d,k+1}(\alpha_1, \ldots, \alpha_d, \alpha_{d+1})
\end{align*}     
for $\alpha_i\in W_i, 1\le i\le d$ and $\alpha_{d+1}\in\mathcal{S}$.
The index $k_1$ requires an explanation. Notice that $k''$ is determined
by $k'$ by the equation $k'=j'\p^{d-2}+k''$. Thus, if $k'$ increases
to $k'+1$ then we have a new equation $k'+1=j_1\p^{d-2}+k_1$ and,
depending on whether $k'< j\p^{d-1}-2$ or not, $k_1=k''+1, j_1=j'$
or $k_1=0, j_1=j'+1$. Either case is covered by the above argument
proving the lemma.\qed

\paragraph{Degree reduction of the generator $\G_d$}

By Remark~\ref{rem:degree} the variable $\alpha_i$ has degree $e_{ik}$
bounded by $\Lambda \ord(\omega_i)-1$ for $1\le i\le d$, and $\alpha_{d+1}$ has degree $r^2 \ord(\omega_d)$, the degree of the interpolating polynomial.  Therefore,
for each $d$ and $0\le k\le \p^d-1$, there is a polynomial $\G'_{d,k}$
over $F$ in variables $\alpha_i, 1\le i\le d+1$ such that
$\deg(\alpha_i)\le \Lambda \p^{\kappa+i} -1$ and $\deg(\alpha_d) =  r^2 \p^{\kappa+d}$. Furthermore, $\G'_{d,k}$ agrees
with the hitting set generator $\G_{d,k}$ on all the points in the
set $W_1\times W_2\times\cdots \times W_d\times \mathcal{S}$ for all $k$.
This defines the reduced degree modified generator $\G'_d$. Following \cref{lem:hsg-span}, the following is immediate:
\begin{equation}\label{eq:modified-hsg}
\Span_K\left\{\prod_{k = 0}^{\p^d-1} M_k(u_k)\right\}_{\text{each~}u_k\in F} \subseteq
\Span_K\left\{\prod_{k=0}^{\p^d-1} M_k(\G'_{d,k}(\alpha_1, \ldots, \alpha_d,
  \alpha_{d+1}))\right\}_{\substack{\alpha_{d+1}\in \mathcal{S}, \\ \text{each~}\alpha_i\in W_i, 1\le i\le d}}.
\end{equation}

\begin{remark}[Construction of the modified generator]\label{rem:roabp_hsg}
  We note that the polynomial $\G'_{d,k}$ has at most
  $O(\Lambda\cdot \p^{2\kappa})^{d+1}$ many monomials. Moreover, its
  values at any point in
  $W_1\times \cdots\times W_d \times \mathcal{S}$ agrees with the
  polynomial $\G_{d,k}$ which can be evaluated at these points in
  parallel time $O((d+1)\cdot \log \Lambda)$.  Thus, finding the
  coefficients of $\G'_{d,k}$ can also be solved in parallel time
  $O((d+1)\cdot \log \Lambda)^{O(1)}$ as solving a system of linear
  equations is in $\NC$. Thus, if $\Lambda$ is quasipolynomial, we have
  a quasi-$\NC$ algorithm for computing the hitting set generator
  polynomial $\G'_d$ in sparse representation.
\end{remark}

\begin{theorem}\label{thm:roabp-hsg}
  For prime $\p$, let $M_k\in F[u_k]^{r\times r}$ for
  $1 \le k \le \p^d$, where each entry of $M_k$ is of degree at most
  $n$. Then the generator $\G'_d:K^{d+1}\to K^{\p^d}$ (whose $k^{th}$
  component $\G'_{d,k}$ is described above) is a hitting set generator
  for the ROABP defined by the matrix product
  $\prod_{k=0}^{\p^d-1}M_k(u_k)$. More precisely, if $f(\ubar{u})$ is
  the polynomial computed at the $(1,1)^{th}$ entry of the matrix
  product $\prod_{k=0}^{\p^d-1}M_k(u_k)$ then $f\equiv 0$ if and only
  if the $(1,1)^{th}$ entry of the matrix product
  $\prod_{k=0}^{\p^d-1}M_k(\G'_{d,k}(\alpha_1,\alpha_2,\ldots,\alpha_{d+1}))=0$
  for all
  $(\alpha_1,\alpha_2,\ldots,\alpha_{d+1})\in W_1\times
  \cdots\times W_d \times \mathcal{S}$.
\end{theorem}

\begin{proof}
  If $f(\ubar{u})$ is nonzero then, as $F$ is an infinite field, we
  can set $u_i=a_i\in F$ such that $f(\ubar{a})\ne 0$. By
  \cref{eq:modified-hsg} there is some
  $(\alpha_1,\alpha_2,\ldots,\alpha_{d+1})\in W_1\times
  W_2\times\cdots \times \mathcal{S}$ such that
  $\prod_{k=0}^{\p^d-1}M_k(\G'_{d,k}(\alpha_1,\alpha_2,\ldots,\alpha_{d+1})\ne
  0$. Conversely, suppose $f\equiv 0$. Then for all
  $a_i\in K, 0\le i\le \p^d-1$ we have $f(\ubar{a})=0$. This implies
  the $(1,1)^{th}$ entry of the matrix product
  $\prod_{k=0}^{\p^d-1}M_k(\G'_{d,k}(\alpha_1,\alpha_2,\ldots,\alpha_{d+1}))=0$
  for all
  $(\alpha_1,\alpha_2,\ldots,\alpha_{d+1})\in W_1\times
  W_2\times\cdots \times \mathcal{S}$.
\end{proof}

\subsubsection{Division algebra hitting set construction}\label{subsubsec:divalg-hsg}
Now we are ready to prove the main theorem of the section, a division algebra hitting set construction for noncommutative ABPs.

\begin{theorem}\label{hitting-set-p-divalg}
  Let $\p>2$ be a prime. For $n$-variate degree-$\tilde{d}$
  noncommutative polynomials computed by homogeneous ABPs of width $r$
  over rationals $\Q$, we construct a hitting set
  $\widehat{\H}_{n,r,\tilde{d}} \subseteq D^n$ of size
  $(nr\tilde{d})^{O(\p \log_{\p} \tilde{d})}$ in
  $(nr\tilde{d})^{O(\p\log_{\p} \tilde{d})}$ time, where $D$ is a cyclic
  division algebra of index $\ell = \p^L$ where $L = O(\log_\p \tilde{d})$.
\end{theorem}

\begin{proof}
  We will set $\ell=\p^L$ as the index of the division algebra $D$,
  where $L$ will be determined in the analysis below. A necessary
  condition is $\p^{L} \ge \tilde{d}$. We assume without loss of
  generality that $\tilde{d}=\p^d$ (and we may increase the degree by
  a factor of at most $\p$).

  A key idea in \cite{FS13} is to convert the given ABP into a
  set-multilinear form and eventually an ROABP. Specifically, they
  replace the noncommutative variable $x_i$ by the matrix $M(x_i)$
  : \[ M(x_i)=
\begin{bmatrix}
     0       & x_{i1} & 0 & \cdots & 0 \\
     0       & 0 & x_{i2} &\cdots  &  0 \\
     \vdots & \vdots &\ddots &\ddots  &  \vdots \\
     0       & 0 &\cdots                  & 0 &x_{i\tilde{d}} \\
     0    & 0 & \cdots & 0 & 0
\end{bmatrix} .
\]
and the (commuting) variables $x_{i1}, x_{i2}, \ldots, x_{i\tilde{d}}$ will in turn be replaced by powers $u^i_1, u^i_2, \ldots, u^i_{\tilde{d}}$ of commuting variables $u_j$.  The variables $u_j$ will be finally substituted in the \cite{FS13} construction by the output of a generator $\mathcal{G}_d$ (where $\p^{d-1}<\tilde{d}\le \p^d$) that stretches a seed $(\alpha_1, \alpha_2, \ldots, \alpha_{d + 1})$ to $(\G_{d,0}(\ubar{\alpha}), \G_{d,1}(\ubar{\alpha}), \ldots, \G_{d,\p^d-1}(\ubar{\alpha}))$.

Obviously, the above matrices are nilpotent and they cannot be
elements of a division algebra. Here, our plan will be to replace
$x_i$ by the following matrix $M(x_i)$ : \[ M(x_i)=\left[
\begin{array}{c c c c c | c c c}
     0       & f^i_0(\ubar{\alpha}) & 0 & \cdots & 0 &0 &\cdots  &0\\     
     0       & 0 & f^i_1(\ubar{\alpha}) & \cdots & 0 &0 &\cdots  &0\\
     \vdots & \vdots &\ddots &\ddots  &  \vdots &\vdots &\ddots &\vdots\\
     0       & 0 & 0 & \cdots & f^i_{\p^d-1}(\ubar{\alpha}) &0 &\cdots  &0\\
     0       & 0 & 0 & \cdots & 0 &f^i_{\p^d}(\ubar{\alpha}) &\cdots  &0\\
     \hline
     \vdots & \vdots &\ddots &\ddots  &  \vdots &\vdots &\ddots &\vdots\\
     0       & 0 & 0 & \cdots & 0 &0 &\cdots  &f^i_{\p^L-2}(\ubar{\alpha})\\     
     zf^i_{\p^L-1}(\ubar{\alpha}) &0 & 0 & \cdots & 0 &0 &\cdots  &0
\end{array}\right], 
\]
where for each $j$, $f_j(\ubar{\alpha}) = \G_{d,j}(\ubar{\alpha})$ is an element from an extension
field $K=F(\omega)$ as described. Furthermore, we will ensure that the
automorphism $\sigma$ of $K$ with $F$ as its fixed field has the
property that $\sigma(f_j(\ubar{\alpha}))=f_{j+1}(\ubar{\alpha})$ and
$\sigma^{\p^L}$ is the identity automorphism. This will ensure that
the matrix substitutions $M(x_i)$ for all choices of $\ubar{\alpha}$
from the hitting set will live in the cyclic division algebra defined
by $(K,F,\sigma)$. But this is precisely achieved by the construction
of the degree-reduced hitting set generator $\G'_d$ in \cref{thm:roabp-hsg}.\\

\noindent\textbf{Size of the hitting set}\\

We first analyze the size of the ROABP hitting set of
\cref{thm:roabp-hsg}.  It is just $|\mathcal{S}|\cdot \prod_{i=1}^{d}|W_i|$. Now
$|W_i|=\Lambda\cdot p^{\kappa+i}$ and $|\mathcal{S}| = \Lambda\p^{k+d}$. Thus
\[ |\mathcal{S}|\cdot \prod_{i=1}^{d}|W_i| \leq \Lambda^{d+1}\cdot
  \p^{\kappa (d+1)}\prod_{i=1}^{d+1}\p^i.
\]
The construction requires $\kappa\ge d+1$. Setting $\kappa=d+1$ gives
us an upper bound of $\Lambda^{d+1}\cdot \p^{2(d+1)^2}$. By the above
construction this is also the size of the hitting set for the
noncommutative ABP. Now, by Lemma~\ref{fs13-lemma-prime}, it suffices
to choose $\Lambda>(\p^dnm)^\p r^2$, where we can see that
$m\le r^2\p^{2d+1}$. Recall that $\Lambda$ is a power of some prime numbber $\mathscr{q}$ (as described in \cref{sec:ROABP-hsg}). Therefore, we can choose $\Lambda \leq \mathscr{q}(\p^dnm)^\p r^2$. We choose $\p$ and $\mathscr{q}$ to be two distinct successive primes, and hence $\mathscr{q} \leq 2\p$. With this choice of $\Lambda$ and $\mathscr{q}$, the hitting set size
is bounded by $2p(\p^{3d+1}nr^2)^\p r^2\cdot \p^{2(d+1)^2}$. Now,
noting that the actual degree satisfies
$\p^{d-1}\le \tilde{d}\le \p^d$ we obtain the size bound
$((\p\tilde{d})^4r^2n)^{\p(\log_\p \tilde{d}+1)}$ which is
$(\tilde{d}rn)^{O(\p\log_\p \tilde{d})}$.
\end{proof}

\subsection{Hitting set for generalized ABPs}

In order to apply \cref{hitting-set-p-divalg} to generalized ABPs, we
basically describe an efficient reduction that computes a hitting set
for generalized ABPs given as input a hitting set for noncommutative
ABPs. This is an important conceptual part of the proof.

{ Informally, the observations contained in the following three
  lemmas, \cref{claim:D1-abp-nonzero},
  \cref{claim:D1-abp-matrix-to-DA}, and \cref{lemma:tensor-existence},
  show that a nonzero polynomial computed by a $D$-ABP can be
  evaluated to nonzero at a point in a cyclic division algebra
  $D\otimes D'$ where the index of the cyclic division algebra
  $D'$ is relatively prime to the index of $D$.

\begin{lemma}\label{claim:D1-abp-nonzero}
For any nonzero $n$-variate degree-$d$ $D$-ABP $\mathcal{A}$, 
for every $d' \geq \ell_1d$, there is a $d'\times d'$ matrix tuple such that the $D$-ABP is nonzero evaluated on that tuple. Here $\ell$ is the index of $D$. 
\end{lemma}

\begin{proof}
  Fix an edge of $\mathcal{A}$ and let its label be
  $\sum_{i=1}^n a_ix_ib_i$, for $a_i, b_i\in D_1$. Replace each
  $a_i, b_i\in D$ by its matrix representation in $\M_{\ell}(K)$
  and the variable $x_i$ by $Z_i$, an $\ell_1\times \ell_1$ matrix
  whose $(j,k)^{th}$ entry is a new noncommuting indeterminate
  $z_{ijk}$. Therefore, each edge is now labeled by an
  $\ell\times \ell$ matrix whose entries are $K$-linear terms in
  $\{z_{ijk}\}$ variables. After the substitution, $\mathcal{A}$ is
  now computing a matrix $M$ of degree-$d$ noncommutative
  polynomials. Clearly, it is an identity-preserving substitution.
I.e.,\ $\mathcal{A}$ is nonzero if and only if $M$ is
nonzero. Therefore, if $\mathcal{A}$ is nonzero, there exists a
$d\times d$ matrix substitution for the $\{z_{ijk}\}$ variables such
that $M$ evaluated on that substitution is nonzero.\footnote{In fact,
  $\lceil d/2\rceil + 1$-dimensional matrix substitutions will suffice
  \cite{AL50}.} Hence, we obtain an $\ell d\times \ell d$ matrix
tuple for the $\ubar{x}$ variables such that $\mathcal{A}$ is nonzero
on that substitution.
\end{proof}

\begin{lemma}\label{claim:D1-abp-matrix-to-DA}
Suppose for a nonzero $n$-variate degree-$d$ $D$-ABP $\mathcal{A}$,
there is a matrix tuple $(p_1, \ldots, p_n)\in \M_{d'}(K)^n$ such that the ABP is nonzero evaluated on that tuple. 
Let $\widetilde{D} = \left({\widetilde{K}}/{F}, \widetilde{\sigma}, z\right)$ be a cyclic division algebra of index $d'$, where $K$ is a subfield of $\widetilde{K}$. Then there is a tuple in $\widetilde{D}^n$ such that the $D$-ABP $\mathcal{A}$ is nonzero evaluated on that tuple as well.
\end{lemma}

\begin{proof}
  Let
  $\left\{\widetilde{C}_{j,k}\right\}_{1\leq j,k\leq d'}$ be
  the basis of the division algebra $\widetilde{D}$ as defined in
  \cref{sec:cyclic}. By \cref{full-space}, we can write each matrix
  $p_i = \sum_{j,k} \lambda_{ijk} \widetilde{C}_{jk}$ where each
  $\lambda_{ijk}\in \widetilde{K}$. Define new commuting
  indeterminates $\{u_{ijk}\}$ and let
  $\widetilde{p}_i = \sum u_{ijk} \widetilde{C}_{jk}$. Evaluating
  $\mathcal{A}$ on $(\widetilde{p}_1, \ldots, \widetilde{p}_n)$ then
  gives a nonzero matrix of commutative polynomials, as it is nonzero
  if $u_{ijk} \leftarrow \lambda_{ijk}$. There is a substitution for
  each $u_{ijk} \leftarrow \gamma_{ijk} \in \Q$ such that such a
  nonzero polynomial evaluates to nonzero. Hence, we can define a
  tuple $(q_1, \ldots, q_n)$ where each
  $q_i = \sum \gamma_{ijk}\widetilde{C}_{jk}$ such that $\mathcal{A}$
  is nonzero on $(q_1, \ldots, q_n)$. Now the proof follows since each
  $q_i \in \widetilde{D}$. 
\end{proof}

\begin{lemma}\label{lemma:tensor-existence}
  For any nonzero polynomial computed by an $n$-variate degree-$d$
  $D$-ABP $\mathcal{A}$,
  there is a cyclic division
  algebra $\widetilde{D}$ of index $\ell\ell'$ (where
  $\ell'\geq d$ and $\ell'$ is relatively prime to $\ell$) and a
  tuple in $\widetilde{D}^n$ such that $\mathcal{A}$ is nonzero
  evaluated on that tuple.
\end{lemma}

\begin{proof}
  Consider a cyclic division algebra $D'$ of index $\ell'$. Define
  $\widetilde{D} = D\otimes D'$. By assumption, $\ell' (\geq d)$ is
  relatively prime to $\ell$. More precisely, suppose
  $D=(K/F,\sigma,z)$ and $D'=(K'/F',\sigma',z')$, where
  $F=\Q(z,\omega_0)$, $K=F(\omega_1)$, $F'=\Q(z',\omega_0')$ and
  $K'=F'(\omega'_1)$, where $\omega_0$, $\omega_0'$, $\omega_1$, and
  $\omega_1'$ are roots of unity whose orders are relatively prime to
  each other. Then we define the field $F_1=\Q(z,\omega_0\omega_0')$,
  and the fields $K_1=F_1(\omega_1)$ and $K_1'=F_1(\omega_1')$ and
  note that the cyclic division algebra $D$ is isomorphic to
  $(K_1/F_1,\hat{\sigma},z)$ and $D'$ is isomorphic to
  $(K_1'/F_1,\hat{\sigma}',z)$ for suitably defined automorphisms
  $\hat{\sigma}$ and $\hat{\sigma}'$. Now by \cref{thm:cyclic-tensor}
  it follows that $\widetilde{D}$ is also a cyclic division algebra of
  index $\ell\ell'$ defined using the field extension
  $F_1(\omega_1\omega_2)$ of $F_1$. The proof follows from
  \cref{claim:D1-abp-nonzero} and \cref{claim:D1-abp-matrix-to-DA}.
\end{proof}
}

The following lemma now reduces the problem of zero testing of a division algebra ABP to a corresponding noncommutaitve ABP.

\begin{lemma}\label{lemma:gen-abp-reduces-to-nc-abp-pit}
  Let $D$ be a cyclic division algebra of index $\ell$ and
  $\mathcal{A}$ be an $n$-variate degree-$d$ $D$-ABP of width $r$ over
  $\{x_1, x_2, \ldots, x_n\}$. Then, there is an $\ell^2n$-variate
  degree-$d$ noncommutative ABP $\mathcal{B}$ of width $\ell r$ such
  that the following holds:
    \begin{itemize}
    \item The polynomial $f$ computed by the $D$-ABP $\mathcal{A}$ is
      nonzero if and only if the polynomial $g$ computed by the
      noncommutative ABP $\mathcal{B}$ is nonzero.

    \item For any matrix tuple $(q_{111}, \ldots, q_{n\ell\ell})$, if
      the polynomial $g$ computed by $\mathcal{B}$ is nonzero on that
      tuple, then the polynomial $f$ computed by the $D$-ABP
      $\mathcal{A}$ evaluated on $(q_1, \ldots, q_n)$ is also nonzero
      where for each $i$, $q_i = \sum_{j,k} C_{jk}\otimes q_{ijk}$ and
      $\left\{C_{jk}\right\}_{1\leq j,k\leq \ell}$ is a basis of
      $D$.
    \end{itemize}
\end{lemma}

\begin{proof}
  Introduce a set of noncommuting indeterminates
  $\left\{y_{ijk}\right\}_{i\in [n], j,k\in [\ell_1]}$.  Consider the
  following mapping:
\[
x_i\mapsto \sum_{j,k} C_{jk}\otimes y_{ijk}.
\]
Equivalently, each $x_i$ is substituted by an $\ell\times \ell$
matrix. Fix a $D$-ABP $\mathcal{A}$. Consider each edge of
$\mathcal{A}$ labeled as $\sum_{i=1}^n a_ix_ib_i$ where
$a_i, b_i\in D$. Replace each $a_i, b_i\in D$ by its matrix
representation in $\M_{\ell}(K)$ and $x_i$ by the
$\ell\times \ell$ matrix $\sum_{j,k} C_{jk}\otimes
y_{ijk}$. Therefore, each edge is now labeled by an
$\ell\times \ell$ matrix whose entries are $K$-linear terms in
$\{y_{ijk}\}$ variables. After the substitution, $\mathcal{A}$ is now
computing a matrix $M$ of degree-$d$ noncommutative polynomials in
$\{y_{ijk}\}$ variables.
\begin{claim}\label{claim:genABP-to-tensor}
If the $D$-ABP $\mathcal{A}(\ubar{x})$ is nonzero then the matrix $M\in \M_{\ell}(\F\angle{\ubar{y}})$ is nonzero. 
\end{claim}
\begin{proof}
If $\mathcal{A}(\ubar{x})$ is nonzero, then it is nonzero evaluated at some $(p_1, \ldots, p_n)\in \widetilde{D}^n$ where $\widetilde{D} = D\otimes D'$ (\cref{lemma:tensor-existence}). We can therefore expand the $D$ component in the $\{C_{jk}\}$ basis and write each $p_i = \sum C_{jk}\otimes q_{ijk}$ for some $q_{ijk}\in D'$. Therefore $M$ is nonzero under the substitution each $y_{ijk} \leftarrow q_{ijk}$.
\end{proof}
We now claim that each entry of $M$ is computable by a small ABP.
\begin{claim}\label{claim:abp-entry}
For each $1\leq j,k \leq \ell_1$, the $(j,k)^{th}$ entry of the matrix $M\in \M_{\ell}(\F\angle{\ubar{y}})$ is computable by an $\ell^2n$-variate degree-$d$ noncommutative homogeneous ABP of width $\ell r$. 
\end{claim}
\begin{proof}
For each vertex $v$ in the $D$-ABP $\mathcal{A}$, make $\ell$ copies of $v$ (including the source $S$ and sink $T$), let us call it $(v,1)$, $\ldots,$ $(v, \ell_1)$. For any two vertices $u$ and $v$, suppose the edge is labeled by $\sum_{i=1}^n a_ix_ib_i$ and $M_{u,v}$ be the corresponding $\ell\times \ell$ matrix after substitution. Then for each $1\leq \hat{j}, \hat{k} \leq \ell$, we add an edge $((u,\hat{j}),(v,\hat{k}))$ labeled by the $(\hat{j}, \hat{k})^{th}$ entry of $M_{u,v}$. Note that product of the edge labels of a path exactly captures the corresponding matrix product. Therefore, if we consider the ABP with source $(S, \hat{j})$ and sink $(T, \hat{k})$, it is computing the  $(\hat{j}, \hat{k})^{th}$ entry of the matrix $M$. Note that the width of the new ABP is $\ell r$.
\end{proof}

We now consider a nonzero entry of the matrix $M$ which is computable by an $\ell^2n$-variate degree-$d$ noncommutative homogeneous ABP of width $\ell r$. 
\end{proof}

\begin{remark}\label{rem:map-depends-on-inclusion-map}
    The map $x_i \mapsto \sum C_{jk}\otimes y_{ijk}$ transforms a $D$-ABP into an element of $\Mat_{\ell}(\F\angle{\ubar{y}})$ where $\ell$ is the index of $D$. Note that, the inclusion map $\iota: \Mat_{\ell}(\F) \to \Mat_{d\ell}(\F)$ defined by $\iota(a) = a\otimes I_d$ is implicitly assumed here (see \cref{sec:prelim}). If a different inclusion map is applied for evaluating the $D$-ABP, the mapping must be modified accordingly. For instance, if the inclusion map is defined by $\iota(a) = I_d\otimes a$, then the mapping becomes $x_i \mapsto \sum y_{ijk}\otimes C_{jk}$.
\end{remark}

We are now ready to prove the main result of this section.

\begin{theorem}[Division algebra hitting set for $D$-ABPs]\label{thm:gen-abp-hs}
    Fix integers $n,r,d$. Let $\widehat{\H}_{n, r, d}$ be a division algebra hitting set for the class of $n$-variate width-$r$ noncommutative ABPs of degree $d$. Then, for any cyclic division algebra $D$ of index $\ell$, we can design a hitting set $\widehat{\H}^D_{n,r,d}\subseteq \widetilde{D}^n$ for the class of $n$-variate  width-$r$ $D$-ABPs of degree $d$ as follows: 
\begin{equation}\label{eq:gen-abp-hs}
    \widehat{\H}^D_{n,r,d} = \left\{ (q_1, \ldots, q_n) : q_i = \sum_{j,k} C_{jk}\otimes q_{ijk}\quad\text{where}\quad (q_{111}, \ldots, q_{n\ell\ell})\in \widehat{\H}_{\ell^2n, \ell r, d} \right\},
  \end{equation}
  where $\widehat{\H}_{\ell^2n, \ell r, d} \subseteq D'^n$ such that $D'$ is a cyclic division algebra of index $\ell'$ which is relatively prime to $\ell$ and $\widetilde{D} = D \otimes D'$ is a cyclic division algebra of index $\ell\ell'$.
\end{theorem}

\begin{proof}
  From the assumption, $\widetilde{D} = D \otimes D'$. Note that, it is a cyclic division
  algebra of index $\ell\ell'$ by \cref{thm:cyclic-tensor}. Our is now to output a division algebra hitting set
  for a $D$-ABP in the cyclic division algebra $\widetilde{D}^n$.
which immediately follows from  \cref{lemma:gen-abp-reduces-to-nc-abp-pit}.
\end{proof}  
  
The above theorem combined with \cref{hitting-set-p-divalg}
immediately yields the following corollary.

\begin{corollary}\label{cor:dabp-hsg}
    Fix two distinct prime numbers $\p$ and $\p'$.
    For the class of polynomials
  computed by $n$-variate $D$-ABPs of degree-$d$ and width $r$,
  where $D$ is a cyclic division algebra of index
  $\ell=\p^{L}$,
  there is an algorithm that outputs a hitting set
  $\widehat{\H}_{\ell^2n, \ell r, d}\subseteq \widetilde{D}^n$ of
  size $(\ell_1 n r d)^{O(\p'\log d)}$  and $\widehat{D}$ is a cyclic
  division algebra of index $\ell\ell'$, for $\ell'=\p'^{L'}$ where
  $L'$ is $O(\log_{\p'} d)$.   
\end{corollary}

\section{Hitting Set for \textsc{Nsingular} Conditioned on a Matrix Tuple}\label{sec:nsing-witness}

Fix a matrix tuple $\ubar{p} = (p_1, \ldots, p_n)$. For any integer $s$, consider the class of linear matrices such that a submatrix of size $s-1$ is invertible on $\ubar{p}$. We say \textit{the class of linear matrices is conditioned on the matrix tuple $\ubar{p}$.}
In this section, we construct a  hitting set for the $\nsing$ problem for the class of linear matrices conditioned on $\ubar{p}$. 

The result of this section is crucial for the hitting set construction for rational formulas in \cref{sec:main-result}. More precisely, we construct the hitting set for rational formulas
inductively on the inversion height. To construct a hitting set for
inversion height $\theta$ from inversion height $\theta-1$, we consider the union of all hitting sets conditioned on a matrix tuple in the hitting set of height $\theta - 1$.


\begin{lemma}\label{lemma:nsing-to-gen-abp-pit}
    Let $D$ be a cyclic division algebra of index $\ell$ and for any $n$, $(p_1, \ldots, p_n)\in D^n$ be a matrix tuple. Fix an integer $s$. Consider an $n$-variate linear matrix $T$ of size $s$, conditioned  on $(p_1, \ldots, p_n)$. Then, there is an $n$-variate degree-$(s-1)$ $D$-ABP $\mathcal{A}$ of width $s - 1$ such that the following holds:
    \begin{itemize}
        \item The linear matrix $T$ is invertible over the free skew field if and only if the $D$-ABP $\mathcal{A}$ is nonzero.

        \item For any integer $\ell'$ and a matrix tuple $(q_1, \ldots, q_n) \in \M_{\ell\ell'}(\Q)^n$, if $\mathcal{A}$ is invertible on that tuple, then for some indeterminate $t$, $T$ evaluated on
        $(tq_1 + p_1\otimes I_{\ell'}, \ldots, tq_n + p_n\otimes I_{\ell'})$ is invertible.
    \end{itemize}
\end{lemma}


\begin{proof}
Let $T(\ubar{p})$ is not invertible.
We can then find two invertible transformations $U, V$ in $\M_s(D)$ 
such that
\[
U \cdot T(p_1, p_2, \ldots, p_n) \cdot  V = 
\left[
\begin{array}{c|c}
I_{s-1}& 0 \\
\hline
0 & 0
\end{array}
\right],
\]
where $I_{s-1}$ is the identity matrix whose diagonal elements  are the identity element of $D$. This is possible since one can do Gaussian elimination over division algebras.  

Notice that $T(\ubar{x} + \ubar{p})=T(\ubar{p}) + T(\ubar{x})$. Hence, we can write
\[
T(\ubar{x}+\ubar{p})=U^{-1}\cdot \left(\left[
\begin{array}{c|c}
I_{s-1}& 0 \\
\hline
0 & 0
\end{array}
\right] + U\cdot T(\ubar{x})\cdot V\right)\cdot V^{-1}. 
\]
Let the invertible submatrix $T'$ of $T$ of size $s-1$ is obtained by removing the $i^{th}$ row and $j^{th}$ column, for some $i,j\in [s]$. We can therefore write,
\[
T(\ubar{x}+\ubar{p})=U^{-1}\cdot
\left[
\begin{array}{c|c}
I_{s-1} - \widetilde{T} & A_j \\
\hline
B_i & c_{ij}
\end{array}
\right]\cdot V^{-1},
\] 
where each entry of $\widetilde{T}, A_j, B_i, c_{ij}$ are  $D$-linear forms in $\ubar{x}$ variables with no constant term. We can simplify it further by multiplying both sides by invertible matrices and writing, 
\begin{equation}\label{eqn:2by2}
T(\ubar{x}+\ubar{p}) =
U^{-1}U'\left[
\begin{array}{c|c}
I_{s-1} - \widetilde{T} & 0 \\
\hline
0 & c_{ij} - B_{i}(I_{s-1} - \widetilde{T})^{-1}A_j
\end{array}
\right] V'V^{-1}. 
\end{equation}

\[
\text{where,~}
U'= \left[
\begin{array}{c|c}
I_{s-1}  & 0 \\
\hline
B_i (I_{s-1} - \widetilde{T})^{-1} & 1
\end{array}
\right]
,
\quad\quad 
V'= \left[
\begin{array}{c|c}
I_{s-1}  & (I_{s-1}-\widetilde{T})^{-1}A_j \\
\hline
0 & 1
\end{array}
\right].
\]
Note that, here $1$ denotes the  unit in the division algebra $D$.
\begin{equation*}
    \text{Let,}\quad P_{ij}(\ubar{x}) = c_{ij} - B_{i}(I_{s-1} - \widetilde{T})^{-1}A_{j}.
\end{equation*}


We can also represent $P_{ij}$ as a formal series: 
\begin{equation}\label{def:D1-series}
P_{ij}(\ubar{x}) = c_{ij} - B_{i}\left(\sum_{k\geq 0}\widetilde{T}^{k}\right) A_{j}.
\end{equation}
This is a generalized series (in $\ubar{x}$ variables) over the division algebra $D$ where the division algebra elements can interleave in between the variables. The following claim truncates this infinite series to a finite value.

\begin{claim}\label{claim:genseries-to-genabp}
Consider a generalized $D$-series $P$ as defined in \cref{def:D1-series}.
\begin{align*}
     {P}(\ubar{x}) &= c - B\left(\sum_{k \geq 0} \widetilde{T}^{k}\right) A,\\
    \text{Define its truncation:}\quad\widetilde{P}(\ubar{x}) &= c - B\left(\sum_{0\leq k\leq s-1} \widetilde{T}^{k}\right) A.\\
   \text{Then } P(\ubar{x})=0 &\Longleftrightarrow \widetilde{P}(\ubar{x})=0.
\end{align*}
\end{claim}

\begin{proof}
    Suppose $P$ is nonzero.
    Substitute each $\{x_i : 1 \leq i \leq n\}$ by the following map used in the proof of \cref{thm:gen-abp-hs}:
    \[
    x_i\mapsto \sum_{j,k} C_{jk}\otimes y_{ijk}.
    \]
    Consider an entry of $\widetilde{T}$ which is of form $\sum_{i=1}^n a_ix_ib_i$ for some $a_i,b_i\in D$.
    Since $C_{jk}, 1\le j,k\le \ell$ is a basis for the division algebra $D$,
    we can write each entry of $\widetilde{T}$ as $\sum \beta_{ijk}\beta'_{ij'k'}C_{jk}x_iC_{j'k'}$ for some $\beta_{ijk}, \beta_{ij'k'}\in F$. 
    Substituting each $x_i$ as above and identifying each $C_{jk}$ with $C_{jk}\otimes 1$, it follows that each entry of $\widetilde{T}$ can be expressed as $\sum_{j,k} (C_{jk}\otimes \sum_i \alpha_{ijk}y_{ijk})$, where each $\alpha_{ijk}\in F$. Therefore, it now computes a series $\sum C_{jk}\otimes f_{jk} \in D\otimes_{F} F\llangle\ubar{y}\rrangle$ . We first observe the following claim. Its proof is omitted as it is a straightforward generalization of the proof of \cref{claim:genABP-to-tensor}.
    
    \begin{claim}
        $P(\ubar{x}) = 0 \Longleftrightarrow \sum C_{jk}\otimes f_{jk} = 0$.
    \end{claim}
    
    Recall that, $D\llangle y \rrangle$ denotes the formal power series in noncommuting $\ubar{y}$ variables where the coefficients are in $D$ and $\ubar{y}$ variables commute with the elements in $D$. We now define the following map:
    \begin{align*}
        \psi: D\otimes_{F} F\llangle y \rrangle &\to D\llangle y \rrangle,\\
        C_{jk}\otimes y_{ijk} &\mapsto C_{jk}y_{ijk}.
    \end{align*}
    Note that, $\psi$ is an isomorphism. Each entry of the matrix $L$ is now of form $\sum_{i,j,k} \gamma_{ijk}y_{ijk}$ (where $\gamma_{ijk}\in D$). Therefore, substituting each $x_i\mapsto \sum_{j,k} C_{jk}\otimes y_{ijk}$ and then applying $\psi$-map on ${P}$ computes a series in $D\llangle y \rrangle$. We can now apply \cref{lem:trucated} and truncate it to degree $s-1$ preserving the nonzeroness.

    Clearly, applying the substitution $x_i\mapsto \sum_{j,k} C_{jk}\otimes y_{ijk}$ and then the $\psi$-map on $\widetilde{P}$ will have the same effect. Therefore, $\widetilde{P}$ is also nonzero.
\end{proof}

\[
\text{We can now write,}\quad
P_{ij}= 0 \Longleftrightarrow \left(c_{ij} = 0 \quad \text{and} \quad\text{for each } 0\leq k \leq (s-1)\ell_{1},\quad B_{i}\widetilde{T}^{k} A_{j} = 0\right),
\]
where each $B_{i}\widetilde{T}^{k} A_{j}$
is a generalized polynomial over $D$, indeed it is a $D$-ABP.

Let $k_0$ be the minimum $k$ such that $B_{i}L^{k} A_{j}\neq 0$. By definition, $k_0 \leq s - 1$. Define, a generalized ABP $\widetilde{P}_{ij} = B_{i}L^{k_0} A_{j}$.
We now reduce the singularity testing of $T$ to identity testing of this $D$-ABP.\footnote{In a recent work \cite{CM23}, a similar idea is used to show a polynomial-time reduction form $\nsing$ to identity testing of noncommutative ABPs in the white-box setting.}


    Let $\widetilde{P}_{ij}$ be zero, therefore $P_{ij}$ is also zero by \cref{claim:genseries-to-genabp}. Following \cref{eqn:2by2}, $T(\ubar{x}+\ubar{p})$, and hence, $T(\ubar{x})$ is not invertible.

    For the other direction, if $\widetilde{P}_{ij}\neq 0$, by assumption there exists a matrix tuple $(q_1, \ldots, q_n)\in \Mat_{\ell\ell'}(\Q)$ 
    such that $\widetilde{P}_{ij}(\ubar{q})$ is invertible. We then evaluate $T(\ubar{x} + \ubar{p})$ on $(tq_1, \ldots, tq_n)$ where $t$ is a commutative variable. Clearly, the infinite series $P_{ij}$ is nonzero at $t\ubar{q}$ since the different degree-$t$ parts do not cancel each other. 
    Also, $(I_{s-1}-\widetilde{T})(t\ubar{q})$ is invertible. Therefore, by definition $T(\ubar{x}+ \ubar{p})$ is also invertible.

It completes the proof.  
\end{proof}

\begin{theorem}\label{thm:nsing-witness}
    Fix integers $n,r,d$. Let $D$ be a cyclic division algebra of index $\ell$ and $\widehat{\H}^{D}_{n, r, d} \subseteq \widehat{D}^n$ be a division algebra hitting set for the class of $n$-variate width-$r$ $D$-ABP of degree $d$ where $\widehat{D}$ is of index $\ell\ell'$. Then, given a matrix tuple $\ubar{p} = (p_1, \ldots, p_n)\in D^n$, we can design a hitting set $\widetilde{\H}^{\ubar{p}}_{n, s, \ell}$ for the class of $n$-variate linear matrices of size $s$ conditioned on tuple $\ubar{p}$ as follows:
    \begin{equation}\label{eq:rank-increment(1)}
    \widetilde{\H}^{\ubar{p}}_{n, s, \ell} = \left\{(aq_1 + p_1\otimes I_{\ell'}, \ldots, aq_n + p_n\otimes I_{\ell'}): \ubar{q}\in \widehat{\H}^{D}_{n, s-1, s-1}\quad \text{and}\quad a\in \Gamma, |\Gamma| = (s\ell\ell')^{O(1)}\right\}.
\end{equation}
\end{theorem}

\begin{proof}
    The proof immediately follows from \cref{lemma:nsing-to-gen-abp-pit} except the bound on $\Gamma$. Note that,
    For any linear matrix $T$ of size $s$, consider the matrix $T$ evaluated on $(tq_1 + p_1\otimes I_{\ell'}, \ldots, tq_n + p_n\otimes I_{\ell'})$. Clearly, it is a matrix of size $s\ell\ell'$  where each entry is a linear polynomial in $t$. Consider the rational expressions in the inverse of this matrix.
    Since the degrees of the polynomials in the rational expressions and the determinant are bounded by a polynomial in the size of the matrix,  and we need to only avoid the roots of the numerator and the denominator polynomials present in each entry, we can vary the parameter $t$ over a set $\Gamma\subset \Q$ such that $|\Gamma| = (s\ell\ell')^{O(1)}$.
\end{proof}

\begin{corollary}\label{cor:nsing-witness}
    Let $D$ be a cyclic division algebra of index $\ell$ and $\p$ be any prime that is not a divisor of $\ell$. Given a tuple $(p_1, \ldots, p_n)\in D^n$, consider the class of $n$-variate linear matrix of size $s$ conditioned on $(p_1, \ldots, p_n)\in D^n$. We can construct a hitting set $\widetilde{\H}^{\ubar{p}}_{n,s, \ell}\subseteq \widetilde{D}^n$ of size $(\ell ns)^{O(\p\log  s)}$ in deterministic $(\ell ns)^{O(\p \log s)}$-time for this class where $\widetilde{D}$ is a cyclic division algebra of index $\ell s^{O(1)}$.
\end{corollary}

\begin{proof}
The proof follows from the combination of \cref{thm:gen-abp-hs} and \cref{thm:nsing-witness}.
\end{proof}

\begin{remark}\label{rem:hs-depends-on-inclusion-map}
    As previously discussed in \cref{rem:map-depends-on-inclusion-map}, the witness matrix tuple is also influenced by the choice of the inclusion map used to evaluate the $D$ division algebra ABP. More precisely, if the inclusion map $\iota: \Mat_{\ell_1}(\F)\to \Mat_{\ell_1\ell_2}(\F)$ is applied for evaluating the $D$-ABP, then the hitting set is redefined as follows:
    \begin{equation*}
    \widetilde{\H}^{\ubar{p}}_{n, s, \ell_{1}} = \left\{(tq_1 + \iota(p_1), \ldots, tq_n + \iota(p_n)): \ubar{q}\in \widehat{\H}^{D}_{n, s-1, s-1}\quad \text{and}\quad a\in \Gamma\right\}.
    \end{equation*}
\end{remark}

\section{Derandomizing Black-box RIT}\label{sec:main-result}

In this section we will first prove \cref{thm:main-theorem} by explaining
the quasipolynomial-size hitting set construction for rational formulas
along with analyzing its size. Then we will outline the quasi-$\NC$ white-box
RIT algorithm for rational formulas.

\subsection{Hitting set construction for rational formulas}\label{sec:final-hs}


The hitting set construction is by induction on the inversion height
of a rational formula. We will show that for every inversion height
$\theta$ we can construct a hitting set $\H_{n,s,\theta}\subseteq
D^n_{\theta}$ as claimed, where $D_{\theta}$ is a cyclic division
algebra of index $\ell_{\theta}$. The base case $\theta=0$ is for
noncommutative formulas (which have inversion height $0$). By
\cref{hitting-set-p-divalg} we have such a hitting set construction of
size $(ns)^{O(\log s)}$ for noncommutative ABPs (and hence for
noncommutative formulas without inversion gates).



Inductively assume that we have such a construction for the class of
$n$-variate rational formulas of size $s$ and inversion height $\theta
-1$.  Let $\Phi(\ubar{x})$ be any rational formula of inversion height
$\theta$ in $\Q\newbrak{\ubar{x}}$ of size $s$. We first show the
following.

\begin{lemma}\label{claim:strong-to-weak}
For every rational formula $\Phi$ of inversion height $\theta$ in
$\Q\newbrak{\ubar{x}}$ of size $s$, there exists a hitting set point
$\ubar{p}\in {\mathcal{H}_{n, s, \theta - 1}}$ at which
$\Phi(\ubar{p})$ is defined.
\end{lemma}

\begin{proof}
Let $\mathcal{F}=\{\mathfrak{g}_i\mid 1\le i\le k\}$ be the subset of
all inverse gates $\mathfrak{g}_i$ in the formula $\Phi$ such that
there are no other inverse gates on the path from the output gate to
the gate $\mathfrak{g}_i$. For each $\mathfrak{g}_i\in \mathcal{F}$,
let $\phi_i$ be the subformula that is input to gate
$\mathfrak{g}_i$. Now, consider the product formula $\phi=\phi_1
\phi_2 \cdots \phi_k$ (where $k= |\mathcal{F}|$). Notice that the
formula $\phi$ is size at most $s$ since for each $i\ne j$, the
subformulas $\phi_i$ and $\phi_j$ are disjoint (and we can account for
the $k-1$ new product gates in $\phi$ for the $k$ gates
$\mathfrak{g}_i$ of $\Phi$). Furthermore, we note that the formula
$\phi$ is of inversion height $\theta-1$. Therefore, as
$\mathcal{H}_{n,s,\theta - 1}$ is a division algebra hitting set for
$\phi$, for some $\ubar{p}\in \mathcal{H}_{n,s,\theta - 1}$,
$\phi(\ubar{p})$ is nonzero and hence invertible. Consequently, each
$\phi_i$ is also invertible at $\ubar{p}$.  As the path from the
output gate of $\Phi$ to each $\mathfrak{g}_i$ has no other inverse
gate, it follows that $\Phi(\ubar{x})$ is defined at $\ubar{p}$.
\end{proof}

If the rational formula $\Phi$ is of size $s$ then, as shown in
\cite[Theorem 2.6]{HW15}, the formula $\Phi$ can be represented as the
top-right corner of the inverse of a linear matrix of size at most
$2s$. More precisely, $\Phi(\ubar{x}) = u^{t} \mathcal{L}^{-1} v$
where $\mathcal{L}$ is a linear matrix of size at most $2s$ and
$u,v\in \Q^{2s}$ are $2s$-dimensional column vectors whose first
(resp. last) entry is $1$ and others are zero. Here $u^t$ denotes the
transpose of $u$.
Therefore, $\Phi^{-1}$ can be written as\cite[Equation 6.3]{HW15}:
\[
\Phi^{-1}(\ubar{x}) =
[
1 ~0 ~\ldots ~0
] 
\cdot 
\widehat{\mathcal{L}}^{-1}
\cdot
\left[
\begin{array}{c}
0\\
0\\
\vdots\\
1
\end{array}
\right]
\quad\quad\text{where}\quad
\widehat{\mathcal{L}} = 
\left[
\begin{array}{c|c}
v  & \mathcal{L} \\
\hline
0 & -u^t
\end{array}
\right].
\] 

By \cref{claim:strong-to-weak} the formula $\Phi$ is defined for some
$\ubar{p}\in {\mathcal{H}_{n, s, \theta - 1}}$. Therefore,
$\mathcal{L}$ is invertible at $\ubar{p}$ (see
\cite[Proposition~7.1]{HW15}).

Our goal is to find a point (that it, a tuple over a suitable division
algebra) such that $\Phi$ evaluates to a nonzero and hence invertible
value in the division algebra. Equivalently, $\Phi^{-1}$ needs to be
invertible at that tuple, and therefore $\widehat{\mathcal{L}}$ is
invertible when evaluated at that tuple \cite[Proposition~7.1]{HW15}.


Note that $\widehat{\mathcal{L}}$ is of size at most $2s +
1$. Moreover, we know a tuple $\ubar{p}\in {\mathcal{H}_{n, s, \theta
    - 1}}$ such that a submatrix $\mathcal{L}$ of
$\widehat{\mathcal{L}}$ of size $2s$ is invertible on $\ubar{p}$. We
can now use the construction of $\widetilde{\H}^{\ubar{p}}_{n, 2s+1,
  \ell_{\theta - 1}}$ (where $\ell_{\theta - 1}$ is the index of the
cyclic division algebra $D_{\theta - 1}$), as described in
\cref{thm:nsing-witness}, to find a tuple $\ubar{q}$ inside a division
algebra of index $\ell_{\theta}$ such that
$\widehat{\mathcal{L}}(\ubar{q})$ is invertible, therefore
$\Phi(\ubar{q})$ is nonzero.

We now obtain the following hitting set $\mathcal{H}_{n , s,\theta}$ for the class of $n$-variate noncommutative rational formulas of height $\theta$ and size $s$ defined by the
three equations below:
    \begin{align}\label{eq:final-hs}
        \mathcal{H}_{n , s,\theta} &= \bigcup_{\ubar{p}\in \mathcal{H}_{n, s,\theta - 1} \subseteq D^n_{\theta - 1}} \widetilde{\H}^{\ubar{p}}_{n, 2s + 1, \ell_{\theta - 1}},\\ 
   \widetilde{\H}^{\ubar{p}}_{n, 2s + 1, \ell_{\theta - 1}} &= \left\{(aq_1 + p_1\otimes I_{\ell}, \ldots, aq_n + p_n\otimes I_{\ell}): \ubar{q}\in \widehat{\H}^{D_{\theta-1}}_{n, 2s, 2s+1}, ~ \ell_{\theta} = \ell_{\theta - 1}\ell',~\text{and} ~ a\in \Gamma\right\}\label{eq2-fh},\\
    \widehat{\H}^{D_{\theta-1}}_{n, 2s, 2s+1} &= \left\{ (q_1, \ldots, q_n) : q_i = \sum_{j,k} C_{jk}\otimes q_{ijk}~:~ (q_{111}, \ldots, q_{n\ell_{\theta - 1}\ell_{\theta - 1}})\in \widehat{\H}_{\ell^2_{\theta - 1}n, 2\ell_{\theta - 1}s, 2s+1} \right\}\label{eq3-fh},
\end{align}

Here $\widetilde{\H}^{\ubar{p}}_{n, 2s + 1, \ell_{\theta - 1}}$ is the hitting set for the class of linear matrices of size $2s+1$ conditioned on a matrix tuple $\ubar{p}\in D^n_{\theta - 1}$ and $\Gamma \subseteq Q$ is a set of size $(s\ell_{\theta - 1}\ell)^{O(1)}$ (\cref{thm:nsing-witness}), $\widehat{\H}^{D_{\theta-1}}_{n, 2s, 2s+1}\subseteq D^n_{\theta}$ is a division algebra hitting set for an $n$-variate width-$2s$ and degree-$2s+1$ $D_{\theta-1}$-ABP (\cref{thm:gen-abp-hs}) and finally $\widehat{\H}_{\ell^2_{\theta - 1}n, 2\ell_{\theta - 1}s, 2s+1}$ is a division algebra hitting set for $\ell^2_{\theta -1 }n$-variate width-$2\ell_{\theta - 1}$ and degree-$2s+1$ noncommutative ABP (\cref{hitting-set-p-divalg}).

Let $\ell$ be the dimension of the matrices in
$\widehat{\H}_{\ell^2_{\theta - 1}n, 2\ell_{\theta - 1}s, 2s+1}$. By
\cref{hitting-set-p-divalg} we obtain $\ell \leq s^{O(1)}$.


Noting that $\ell_\theta=\ell\cdot \ell_{\theta -1}$, we now have $\ell_\theta = s^{O(\theta)}$.

We also have from \cref{eq:final-hs}, \cref{eq2-fh}, \cref{eq3-fh} that, 
\[
|\H_{n , s,\theta}| = |\H_{n , s,\theta - 1}|\cdot |\widehat{\H}_{\ell^2_{\theta - 1}n, 2\ell_{\theta - 1}s, 2s+1}|\cdot |\Gamma|.
\]

We now apply
\cref{thm:nsing-witness} to bound $|\Gamma|$ by $(s\ell\ell_{\theta
  -1})^{O(1)}$. Therefore, $|\Gamma| = s^{O(\theta)}$. Let $\p_{\theta}$ be the $\theta^{th}$ prime number. In order to apply
\cref{hitting-set-p-divalg}, and consequently \cref{thm:gen-abp-hs}, we choose $\p_{\theta+1}$ 
and $\p_{\theta+2}$ in the construction of $\widehat{\H}_{\ell^2_{\theta - 1}n, 2\ell_{\theta - 1}s, 2s+1}$. As evident in the proof of \cref{hitting-set-p-divalg}, the size of the hitting set is bounded by $(ns^2\ell^3_{\theta - 1})^{O(\theta^2 \log s)}$ as $\p_{\theta}\leq \theta^2$.
\begin{align*}
    |\widehat{\H}_{\ell^2_{\theta - 1}n, 2\ell_{\theta - 1}s, 2s+1}|\cdot |\Gamma| &\leq 
    (ns^2\ell^3_{\theta - 1})^{O(\theta^2 \log s)} \cdot s^{O(\theta)}\leq (ns\ell_{\theta - 1})^{O(\theta^2\log s)}.
\end{align*}
As shown above $\ell_{\theta - 1}\le s^{O(\theta)}$. Hence, we now obtain
\[
|\H_{n , s,\theta}| = |\H_{n , s,\theta - 1}|\cdot (ns\ell_{\theta - 1})^{O(\theta^2 \log s)} \leq |\H_{n , s,\theta - 1}| (ns)^{O(\theta^3 \log s)}.
\]
Therefore, $|\H_{n , s,\theta}| \leq (ns)^{O(\theta^5 \log s)}$.

\paragraph{Final steps}


Note that $\H_{n,s,\theta} \subseteq D^n_{\theta}$. From our
construction, the entry of each matrix in the hitting set is in
$\Q(z, \ubar{\omega_0}, \ubar{\omega})$ where $z$ is a commuting
indeterminate;
$\ubar{\omega_0} = (\omega_{0,1}, \ldots, \omega_{0, \theta})$ where
each $\omega_{0,i}$ is a root of unity of order
$\Lambda_i = \mathscr{q}^\tau_i$; and
$\ubar{\omega} = (\omega_{1}, \ldots, \omega_{\theta})$ where each
$\omega_{i}$ is a root of unity of order $\mathscr{p}^\tau_i$, and all
$\{\p_{i}\}_{0 \leq i \leq \theta}$ and
$\{\mathscr{q}_{i}\}_{0 \leq i \leq \theta}$ are distinct primes.

Firstly, we need to bound the parameter $\Lambda_\theta$ for each
inversion height $\theta$. As described in the construction of
\cref{hitting-set-p-divalg}, $\Lambda_\theta = (ns\ell_{\theta -
  1})^{O(\theta^2)}$. From the bound on $\ell_\theta$, therefore,
$\Lambda_\theta = (ns)^{O(\theta^3)}$.  Note that, the degree of each
$\omega_i$ and of $z$ is of degree $\leq s^{O(\theta)}$.

We now discuss how to obtain a hitting set over $\Q$ itself. In the
hitting set points we replace (as discussed in Section
\ref{sec:proofidea}) $\ubar{\omega}$ and $\ubar{\omega_0}$ by
commuting indeterminates $\ubar{t_1}, \ubar{t_2}$.  Then, for any
nonzero rational formula $\Phi$ of size $s$ there is a matrix tuple in
the hitting set on which $\Phi$ evaluates to a nonzero matrix
$M(z, \ubar{t_1},\ubar{t_2})$ of dimension $s^{O(\theta)}$ over the
commutative function field $\Q(z,\ubar{t_1},\ubar{t_2})$. Each entry
of $M(z, \ubar{t_1},\ubar{t_2})$ is a commutative rational expression
of the form $a/b$, where $a$ and $b$ are polynomials in $z$,
$\ubar{t_1}$ and $\ubar{t_2}$ and the degrees of both $a$ and $b$ are
$(ns)^{O(\theta^3)}$. We can now vary the variables 
$z, \ubar{t_1}, \ubar{t_2}$ over a $(ns)^{c\theta^3}$-size large set
$\widetilde{T}\subseteq \Q$ for some sufficiently large constant $c$
such that it avoids the roots of all the numerator and denominator
polynomials involved in the computation.  Therefore, finally we obtain
the hitting set $\H_{n,s,\theta}\subseteq \M_{\ell_{\theta}}(\Q)$
(with a slight abuse of notation) where
\[
\ell_{\theta}\leq s^{O(\theta)}\quad\quad\text{and,}\quad |\H_{n,s,\theta}|\leq (ns)^{O(\theta^4 \log s)}.
\]
It completes the proof of \cref{thm:main-theorem}.

\subsection{RIT is in quasi-NC}\label{sec:quasinc}

Recall that, $\NC$ is the class of problems which can be solved in
poly-logarithmic time using polynomially many processors in
parallel. Similarly, quasi-$\NC$ is the class of problems which can be
solved in poly-logarithmic time using quasipolynomially many
processors in parallel. We now prove \cref{thm:main-theorem1}, by
outlining a quasi-$\NC$ RIT algorithm for rational formulas in the
white-box setting. The proof consists of two steps. Firstly we show
that the hitting set presented in the last section can be constructed
in quasi-$\NC$. We then show that given a matrix tuple, a
noncommutative rational formula can be evaluated in $\NC$. We now
describe each step.

\begin{description}
    \item[Step~1. Quasi-$\NC$ hitting set construction:] Firstly, note
      that the matrix operations like additions, and tensor products
      are routinely computable in $\NC$. The hitting set generator
      (see \cref{eq:generator-definition}) requires repeated
      application of $\sigma$ to a root of unity, say $b$, in order to
      compute the sequence $\sigma^i(b), 0\le i\le \p^L-1$. However,
      as the automorphism $\sigma$ is defined by exponentiation, we
      can compute all these $\sigma^i(b)$ in parallel by
      exponentation. Now, for each $i$ we will need to compute a power
      $b^N$ for some $N$ which can be done in $\log N$ parallel rounds
      by repeated squaring. Moreover, for the construction described
      in \cref{eq:generator-definition}, the size of $N$ in binary is
      bounded by quasipolynomial in the input size. Hence this
      computation too is in quasi-$\NC$. Putting it together, as
      argued in \cref{rem:roabp_hsg}, the division algebra hitting set
      construction for the noncommutative ABPs is in quasi-NC.

      Now, from the description of the hitting set $\H_{n, s, \theta}$
      given in \cref{sec:final-hs} via \cref{eq:final-hs},
      \cref{eq2-fh}, \cref{eq3-fh}, it is easy
      build the hitting set in quasi-$\NC$ by induction on $\theta$.
      It is directly based on $\theta$ many constructions of division
      algebra hitting sets for noncommutative ABPs (with different
      size and width parameters at each stage). Each of these $\theta$
      many computations is in quasi-NC as we outlined above and
      $\theta$ itself is bounded by $\log s$.

      
\end{description}

\paragraph{Step 2. Parallel evaluation of rational formulas:}\

Let $\Phi(x_1,x_2,\ldots,x_n)$ be a rational formula of size $s$ in
the noncommuting variables $x_i$. Given a matrix tuple
$(p_1,p_2,\ldots,p_n)$ of $\ell\times \ell$ matrices over $\Q$, our aim is
to give an $\NC$ algorithm for evaluating $\Phi(p_1,p_2,\ldots,p_n)$
if $\Phi$ is defined at this matrix tuple and otherwise detecting
that it is undefined.

We first note that if the formula $\Phi$ has depth $O(\log s)$ then it
is amenable to parallel evaluation on the input $(p_1,p_2,\ldots,p_n)$
using the formula structure. Matrix multiplication, addition, and
matrix inversion are all in $\NC^2$ \cite{Csanky76, Berko84}. Hence evaluation of
$\Phi(p_1,p_2,\ldots,p_n)$ is in $\NC^3$ in this case.

In general, the formula $\Phi$ may have depth $O(s)$. Hrubes and
Wigderson, in \cite[Proposition 4.1]{HW15}, have shown that any
noncommutative rational formula of size $s$ can be transformed into an
equivalent rational formula of depth $O(\log s)$ (and hence size
$s^{O(1)}$). Their transformation, though algorithmic, uses RIT for
rational formulas as a subroutine.\footnote{As observed
in \cite{AJ24}, depth-reduction of noncommutative rational formulas is
$\NC$-reducible to RIT for rational formulas.} Now, as RIT for
rational formulas is in deterministic polynomial
time \cite{IQS18,GGOW16}, it follows that rational formula
depth-reduction is in deterministic polynomial time. 


However, our aim is only \emph{parallel evaluation} of a rational
formula of size $s$ (without any depth constraint) at a matrix tuple
$(p_1,p_2,\ldots,p_n)$.  Examining the Hrubes and Wigderson,
in \cite[Proposition 4.1]{HW15} depth reduction of rational formulas,
we note that that their algorithm is essentially based on Brent's
classical result \cite{Brent74} on depth reduction for commutative
arithmetic formulas.\footnote{We note that Brent actually describes a
detailed parallel algorithm for carrying out the depth reduction.}
However, there are some new aspects. It turns out that if $\Psi$ is a
rational formula in noncommuting variables $z, y_1,y_2,\ldots, y_m$
with $z$ occurring exactly once as input then $\Psi$ has a $z$-normal
form expression:
\[
\Psi=(Az+B)(Cz+D)^{-1}
\]
where $A, B, C, D$ are small rational formulas with no occurrence of
$z$. They exploit this structure in their divide and conquer algorithm
for constructing the equivalent formula $\widehat{\Phi}$ in polynomial
time.

Using this structure, there is a simple $\NC$ algorithm \cite{Jog23}
for the evaluation of $\Phi(p_1,p_2,\ldots,p_n)$, given as input a
rational formula $\Phi$ and a matrix tuple $(p_1,p_2,\ldots,p_n)$
which we briefly describe below.

\begin{enumerate}
\item The input rational formula $\Phi$ is a binary tree. Let $r$
  denote its root. By standard $\NC$ computation we can find a gate
  $v$ in $\Phi$ such that the size of the subformula $\Phi_v$ rooted
  at $v$ has size between $s/3$ and $2s/3$.
\item We compute the path $P=(v,v_1,v_2,\ldots,v_t=r)$ of all gates
  from $v$ to $r$ in $\Phi$.  Then we find all the gates $u$ in $\Phi$
  such that $u\notin P$ and $u$ is input to some gate $v_i\in
  P$. Notice that for $v_i\in P$ such that $v_i\in\{+,\times\}$ has
  exactly one such input $u$. The inversion gates $v_i$ are unary.
\item Recursively evaluate $\Phi_v$ and each such $\Phi_u$ on the
  input $(p_1,p_2,\ldots, p_n)$.
\item We are left with the problem of evaluation a \emph{skew}
  rational formula $\Phi'$ consisting of the gates along path $P$ with
  the already computed $\Phi_u$ and $\Phi_v$ as inputs. Using
  $z$-normal forms \cite{HW15} (defined above) it is easy to obtain a
  simple divide-and-conquer parallel algorithm for evaluating skew
  rational formulas (in particular, evaluating $\Phi'$) (as described
  in the claim below).
\end{enumerate}

More precisely, a \emph{skew rational formula} $\Phi'$ is a rational
formula whose internal gates $v_1,v_2,\ldots,v_t=r$, where $r$ is the
output gate, form a path. Every gate $v_i, i>1$ that is a $+$ or
$\times$ has one input that is a formula input. That is, in the binary
tree underlying the skew formula if $v_i, i>1$ has two children then
exactly one of them is a leaf node. If the gate $v_1$ is an inversion
then it has a single input (which is leaf). Otherwise,
$v_1\in\{+,\times\}$ will have two leaves as children.

\begin{claim}
Let $\Phi'$ be a skew rational formula with internal
$v_1,v_2,\ldots,v_t=r$, where $r$ is the output gate. Then there is an
$\NC^3$ algorithm to evaluate $\Phi'$ on a given matrix assignment to
the input variables.
\end{claim}

\begin{claimproof}
Every non-inversion gate $v_i$ for $i>1$ in $\Phi'$ is either a $+$ or
a $\times$ and has one input that is a matrix (over $F$).

We will solve a more general problem where the last gate $v_1$ in
$\Phi'$ is variable $z$. Then, using the $z$-normal form of
\cite{HW15}, for a matrix assignment to all input variables the
formula $\Phi'$ evaluates to $(Az+B)(Cz+D)^{-1}$, where $A, B, C, D$
are matrices over $F$. We will show that this $z$-normal form is
computable in $O(\log t)$ many parallel rounds of matrix operations
(which are matrix multplications and inversions which can be performed
in $\NC^2$ using standard algorithms).

We split $\Phi'$ into two skew formulas: formula $\Phi'_1$ defined by
the path $z=v_1,v_2,\ldots,v_{t/2}$ and $\Phi'_2$ defined by the path
$v_{t/2}=z_1,\ldots,v_t=r$. Recursively suppose now that we have
computed for the given matrix input $\Phi'_1=(A_1z_1+B_1)(C_1z_1+D_1)^{-1}$ and
$\Phi'_2=(A_2z+B_2)(C_2z+D_1)^{-1}$, where
$A_1,B_1,C_1,D_1,A_2,B_2,C_2,D_2$ are matrices over $F$. Then, exactly as
shown in the proof of \cite[Proposition 4.1]{HW15}, we can write
$\Phi'=(A_1h_1+B_1h_2)(C_1h_1+D_1h_2)^{-1}$, where $h_1=A_2z+B_2$ and
$h_2=C_2z+D_2$. This means we can recover the $z$-normal form for
$\Phi'$ in $\NC^2$ with some matrix operations. This sets up a
divide-and-conquer algorithm on the path length $t$ which means
we have an $\NC$ algorithm that has at most $\log t\le \log s$ parallel
rounds of $\NC^2$ computations giving us an $\NC^3$ upper bound. \qed
 
\end{claimproof}

We analyze the running time of the above parallel algorithm. Let
$T(s)$ bound the number of rounds of parallel matrix multiplications,
additions, inversions required to evaluate a size $s$ rational
formula. Then, the above algorithm yields the bound
\[
T(s)\le T(2s/3) + O(\log s),
\]
which implies $T(s)\le O(\log^2 s)$. Notice that the term $T(2s/3)$
bounds the running time for recursive evaluation of $\Phi_v$ and each
$\Phi_u$, all in parallel, because each of these subformulas have size
at most $2s/3$. The term $O(\log s)$ is the bound\footnote{Here again
we mean $O(\log s)$ rounds of parallel matrix inversions,
multiplications, and additions.} for the separate parallel algorithm,
mentioned above, for evaluating the \emph{skew} rational formula
$\Phi'$.\footnote{A skew rational formula is a rational formula in
which at least one input to each binary gate is a formula input.}

As each matrix operation can be performed in $\NC^2$, it follows that
rational formula evaluation is in $\NC^4$. We obtain the following.

\begin{lemma}
  There is an $\NC^4$ algorithm for evaluating a noncommutative
  rational formula $\Phi$ on a given matrix input
  $(p_1,p_2,\ldots,p_n)$.
\end{lemma}

\begin{remark}
  Notice that the $\NC$ algorithm described above, with minor changes,
  will yield an $O(\log^2 s)$ depth, $\poly(s)$ size rational formula
  equivalent to $\Phi$.
\end{remark}

The size of our final hitting set is
$(ns)^{O(\theta^5\log s)}=(ns)^{O(\log^6 s)}$ and the dimension
of the matrices in the hitting set is
$s^{O(\theta)}=s^{O(\log s)}$. Using the rational formula
evaluation procedure, on each such matrix tuple, it can be evaluated
within quasi-$\NC$. This is in parallel repeated for
$(ns)^{O(\log^6 s)}$ points in the hitting set. This completes the
proof of \cref{thm:main-theorem1}.

\section{Conclusion}\label{sec:conclusion}

In this paper, we nearly settle the black-box complexity of the RIT
problem.  However, designing a black-box algorithm for the $\nsing$
problem remains wide open. The connection of this problem to the
parallel algorithm for bipartite matching \cite{FGT21} is already
discussed in \cref{sec:intro}.

We believe that the techniques introduced in this paper might be
useful in designing efficient hitting sets for the $\nsing$ problem.

Recall that, the result of Derksen and Makam \cite{DM17} implies that
for a nonzero rational formula of size $s$, there is a $2s\times 2s$
matrix tuple such that the evaluation is nonzero. Therefore, the
quasipolynomial bound on the dimension of the hitting set point
obtained in \cref{thm:main-theorem} is far from the optimal bound
known. An interesting open problem is to construct a hitting set where
the dimension is polynomially bounded in the size of the formula.

Another interesting problem is to show that, in the white-box setting
$\rit$ can be solved in $\NC$. Recall that, the identity testing of
noncommutative formulas can be performed in $\NC$ in the white-box
setting \cite{AJS09, Forbes14}.\\

\noindent\textbf{Acknowledgments.}~~We are grateful to the referees
for their several thoughtful comments and suggestions which have helped
us organize the paper and results. We thank Pushkar S. Joglekar for
discussions concerning depth reduction of rational formulas in
parallel, and for sharing his observation~\cite{Jog23}. We also thank
the reviewers of the conference version of this work for their helpful
comments.

\bibliographystyle{alpha}
\bibliography{rit_jnl_revision}
\end{document}